%Paper: astro-ph/9511118
%From: ponomarev@rasfian.serpukhov.su
%Date: Sun, 26 Nov 1995 05:35:45 +0300

\magnification=\magstep0
\tolerance=1000
\hsize=16.0truecm
\vsize=23.5truecm
\topskip=1truecm
\raggedbottom
\abovedisplayskip=3mm
\belowdisplayskip=3mm
\abovedisplayshortskip=0mm
\belowdisplayshortskip=2mm%
\def\al{$\alpha$}
\def\be{$\beta$}

\def\ls{\vskip 12pt}
\def\et{{\it et\thinspace al.}}
\def\mwc {MWC349}
\def\gapprox{$_>\atop{^\sim}$}
\def\lapprox{$_<\atop{^\sim}$}
\def\kms {km\thinspace s$^{-1}$}
\def\ref#1 {\noindent \hangindent=24.0pt \hangafter=1 {#1} \par}   \def\vol#1
{{\bf {#1}{\rm,}\
}}
\def\apj   {{\it Ap.~J.~{\rm,}\ }}
\def\apjl  {{\it Ap.~J.~(Letters){\rm,}\ }}
\def\apjs  {{\it Ap.~J.~Suppl.{\rm,}\ }}

\def\aap   {{\it Astr.~Ap.{\rm,}\ }}
\def\aas   {{\it Astr.~Ap.~Suppl.~Ser.{\rm,}\ }}
\def\anrev {{\it Ann.~Rev.~Astr.~Ap.{\rm,}\ }}

\def\mnras {{\it M.N.R.A.S.{\rm,}\ }}

\def\ats   {{\it Astron.~Tsirk.~{\rm,}\ }}

\def\paz   {{\it Pis'ma v Astron.~Zh.~{\rm,}\ }}

%
%example:
%
%  \ref{Hoffman, D. 1987, \apj\vol{334} 234.}
%
%\def\mo    {M_$\odot$}
%
%
%
\centerline{HYDROGEN MASERS. I: THEORY AND PROSPECTS}
\ls
\ls
\centerline{Vladimir S. Strelnitski $^{1,2}$,
Victor O. Ponomarev $^3$, and Howard A. Smith $^1$}
\vskip 48pt
\ls
\ls
\ls
$^1$ Laboratory for Astrophysics, National Air and Space Museum,
Smithsonian Institution,

Washington, DC 20560

$^2$ New Mexico Institute of Mining and Technology, Socorro, NM 87801

$^3$ P.N. Lebedev Physical Institute, 117924 Moscow, Leninsky
prospect 53, Russia
\vskip 96pt
\ls
\centerline{Submitted to the {\it Astrophysical Journal, Part 1}}
\vfill\eject
\centerline{ABSTRACT}
\vskip 6pt
The discovery of the first naturally occurring high gain hydrogen
recombination line (HRL)
maser, in
the  millimeter and submillimeter spectrum of the emission line  star \mwc,
requires an expansion of current paradigms about HRLs. In this
paper we re-examine in general the physics of non-LTE populations in
recombining hydrogen and specify the conditions necessary for high-gain
masing and lasing in HRLs. To do so we
use the extensive new results on hydrogen level populations produced by
Storey and
Hummer (1995), and our calculations for the net (that is, line plus
continuum) absorption coefficient for the hydrogen, and we present results
for the \al- and \be-lines whose   principal quantum numbers $n$ are
between 5 and 100, for gas whose electron number density
$3 \le {\rm log}\,N_e\,({\rm cm^{-3}}) \le 11$, at
two   electron temperatures, $T_e = 5,000\,$ and 10,000 K. We show
that the unsaturated maser gain in an HRL is a
sharp function of
$N_e$, and thus to achieve
high-gain masing each line requires a sufficiently extended   region over
which the density is rather closely specified.

Saturation of masing recombination lines is a critical
consideration. We derive a simple equation  for estimating the degree   of
saturation from the observed flux density
and the interferometric and/or model information about the   amplification
path length, avoiding the vague issue of the solid   angle of masing.   We
also present a qualitative way to approach the effects of saturation on
adjacent emission lines, although the detailed modeling is highly
case-specific.

We draw attention to another non-LTE phenomenon active in hydrogen:
the
{\it overcooling} of populations.  This occurs for HRLs with
$n$\gapprox 20, in gas where $N_e$\lapprox $10^5\, {\rm cm^{-3}}$.
Observationally the HRL overcooling
might manifest itself as an
anomalously weak emission recombination line, or as a ``dasar,'' that is, an
anomalously strong absorption line. In the simplest case of a   homogeneous
HII region the absorption can be observed on the proper   free-free
continuum of the region, if some conditions for the line   or/and continuum
optical depths are satisfied.

We briefly discuss the prospects of detecting hydrogen masers,
lasers and dasars in several classes of galactic and extragalactic
objects, including compact HII regions, Be or Wolf-Rayet stars, and AGNs.
\ls
{\it Subject headings:} masers --- hydrogen: recombination lines --- stars:
WR stars --- stars: Be
stars --- galaxies: AGN ---stars:particular:MWC 349.
\vfill\eject
\centerline{1. INTRODUCTION}
\ls
Masing spectral lines are an efficient observational  probe in astrophysics:
because of their
strength and narrowness they often provide the best opportunity for
detailed studies of kinematics
and  structure of the emitting source.
Until recently only {\it molecular}
astrophysical masers have been known.
Yet calculations of the atomic hydrogen level populations in HII
regions
which have been done since the 1930's in order to interpret optical
hydrogen recombination lines (HRLs),
showed that {\it global population inversions} probably exist across the
Rydberg levels, thus raising
in principle the possibility of lasing and masing in the HRLs.
Numerical results directly implying inversion   are found in papers like
Cilli\'e (1936), Baker and Menzel (1938), and, later, in Seaton's series of
papers starting in 1959.

In spite of that, and also in spite of the fact that Menzel (1937) pointed out
(even before  the discovery of laboratory
masers and lasers) that under non-LTE  conditions the integral of energy
absorbed by a line may in principle become negative, the   possibility of
lasing in the  HRLs attracted no
attention for many years. It was  perhaps due partly to the deeply rooted
presumptions that the correction factor for stimulated emission, $1 - {\rm
exp}(-h\nu/kT)\,$, is
negligible for optical lines because of their high frequency, and that HII
regions are always
optically thin except perhaps in the Lyman HRLs, so that no serious
amplification effects are possible anyway. Only relatively recently did these
assumptions change. After modeling of optical HRLs in extragalactic objects
suggested HII regions of unusually high densities (up to $N_e \sim
10^{10}\,$cm$^{-3}$) and/or large extensions, Krolik and McKee (1978)
considered the theoretical possibility
of lasing in HRLs.  Smith \et\ (1979), analyzing their observations   of an
unusually intense infrared HRL in
a dense HII region around the young stellar object BN in Orion,   proposed
that if the inversions
persisted to the high densities expected in the source they could   lead to
laser amplification of this line.

In the mid-1960's radio recombination lines were discovered.  For them the
correction factor for
stimulated emission is much more important than  for the optical and IR
lines. Goldberg (1966)
showed that even in HII regions having optical thickness $|\tau|< 1$ in both a
radio recombination line and the adjacent free-free continuum, the
inversion of
populations can increase the line-to-continuum intensity ratio by as much as
a factor of a few.

The first {\it high gain} HRL maser was discovered in the millimeter
recombination lines in the emission line star \mwc\ (Martin-Pintado   \et\
1989a). At first the maser emission was ascribed to the ionized   outflow
from the star (Martin-Pintado \et\ 1989a,b), but later it was
partly (Ponomarev \et\ 1989), and then fully, attributed to the {\it
circumstellar disk} (Gordon 1992; Planesas \et\ 1992; Thum \et\   1992).
Subsequent observational and interpretational papers  enlarged on    the list
of observed masing
lines in this source, and elaborated on the model in terms of a   partly
ionized masing Keplerian disk
[Martin-Pintado \et\ 1989b; Thum   \et\ 1994a,b; 1995; Gordon 1994;
Ponomarev \et\ 1994 (``Paper 1''); Strelnitski \et\ 1995a; 1995b   (``Paper
3'')].

With the discovery of real, high gain, HRL masers in \mwc,   many basic
theoretical
questions arise about the physics of the pumping mechanism that
maintains the inversion, the
dependence of maser gain on the physical parameters (density,
temperature, gas velocity) of the recombining
gas, and about other possible non-LTE  phenomena in HRLs. Directly
related to these is the more practical question of discovering these non-LTE
phenomena in other lines and in other sources. Though briefly touched
upon in some papers (e.g.: Walmsley  1990, Ponomarev \et\ 1991), these
questions deserve to be studied on a more systematic basis.

In this paper we readdress the theory of non-LTE phenomena in HRLs.
Using the
nomenclature for the non-LTE states of a quantum  transition originally
proposed by one of us
(Strelnitski, 1983), along with  the recent and extensive calculations of
hydrogen level populations by
Storey and Hummer (1995), we analyze this phenomena (Section 2).  We
demonstrate   that the
population calculations predict not only inversion and  its weaker
complement, ``overheating'',  both of which
increase the intensities of emission lines, but also their opposite ---
``overcooling''   of
populations --- which tends to produce an anomalously strong absorption   in
the line (the ``dasar''
effect). We briefly discuss from a physical and thermodynamical   point of
view how these
non-LTE states are created in recombining hydrogen.

We address in Section~3  the conditions necessary for producing high
maser   gain in an HRL,  and the conditions (and
observational criteria) of {\it saturation} .
The ``net'' absorption coefficient
which accounts for both the negative absorption (amplification)
in the line and the
positive free-free absorption at the line frequencies, proved to be a
useful parameter (Ponomarev 1994).  Our calculations of the net absorption
coefficient allow us to derive the
optimum electron density $N_e$ for strong unsaturated masing in \al-
and \be-lines with $5 \le n \le  100$
for gas with densities up to $N_e = 10^{12}\,{\rm cm^{-3}}$, and at
two representative values of electron  temperature, $T_e = 5,000$   and
10,000~K. The
results are illustrated graphically and tabulated. Rough analytical
approximations are provided for the lower $n$ transitions. We derive a
formula  for estimating the degree of saturation in an observed masing
HRL, if the line is formed at its optimum electron density.

Section~4 is devoted to the phenomenon of overcooling and ``dasing''   in
HRLs. We use   the
solution of the line-plus-continuum radiative transfer equation   for   a
homogenous slab to show
that direct evidence of overcooling   of a line would be its  appearance in
absorption on the  free-free continuum
of the HII region. We derive the minimal optical depths in the line   and/or
the continuum necessary for this to happen. This is not,   however,  the only
way to detect overcooling: the increment of HRLs   seen in emission can
demonstrate it as well.

Finally, in  Section~5, we discuss  the prospects of detecting HRL
masers,
lasers and/or dasers in galactic and extragalactic sources other than \mwc.
\mwc\ itself --- the only high-gain HRL maser detected so far --- is
addressed separately in a
forthcoming paper (Strelnitski \et\ 1995$b$, ``Paper 3'').
\ls
\ls
\centerline {2. NON-EQUILIBRIUM STATES OF RYDBERG
TRANSITIONS}
\centerline {IN ATOMIC HYDROGEN }
\ls
\centerline {\it {2.1. Radiative Transfer}}
\ls
Our analysis of non-equilibrium effects in HRLs is based on the   standard
solution of the
radiative transfer equation for the   simplest case of a steady-state,
homogeneous, plane-parallel
medium of  finite optical thickness $\tau_{\nu}$:
$$
I_{\nu} = S_{\nu}\, [1 - {\rm exp}(-\tau_{\nu})]\;. \eqno (2.1.1)
$$
Here $I_{\nu}$ is the outgoing radiation intensity at the frequency $\nu$
and $S_{\nu}$ is
the source function. No external background source is assumed.
In the case of the HRLs, where both line and free-free continuum are formed
together, it is convinient to represent the sorce function as $S_{\nu} =
\eta_{\nu} B_{\nu}(T_e)\,$ (Goldberg 1966), where
$$
\eta_{\nu} \equiv {{S_{\nu}} \over {B_{\nu}(T_e)}} = {{k_c' +  k_l'b_2} \over
{k_c' + k_l' b_1 \beta_{12}}}\;; \eqno (2.1.2)
$$
$B_{\nu}(T_e)\;$ is the Planck function for the electron temperature
$T_e$; $k_c'\;$ and
$k_l'\;$ are the continuum and the line   absorption coefficients at   LTE,
both corrected for
stimulated  emission (the prime indicates  LTE values);   $b_n\;$ is   the
Menzel's coefficient of
departure of the   level $n$ population from its LTE value.
$\beta_{12}\,$ in Eq.~(2.1.2) is formally
the ratio of the actual (non-LTE) and the LTE
correction factors for stimulated emission   (Brocklehurst and Seaton
1972):
$$
\beta_{12} = {{1 - (b_2/b_1)\, {\rm exp}\,(-h\nu_0/kT_e)} \over {1 - {\rm
exp}\,(-h\nu_0/kT_e)}} \;, \eqno (2.1.3)
$$
where $\nu_0$ is the rest frequency of the 1-2 transition and constants
have their usual meanings.
$\beta_{12}\,$ is a measure of the deviation of the
population {\it ratio} of two levels from its LTE value.

The net optical depth at the line center  for a homogeneous medium is   (we
omit the subindex $\nu_0$):
$$
\tau_{net} = \tau_c + \tau_l = k_c'L + k_l'\,b_1 \beta_{12} L \equiv (k_c'
+ k_l)\,L \equiv k_{net}L \;. \eqno (2.1.4)
$$
where $k_l \equiv k_l'\,b_1 \beta_{12}\,$ is the
actual   (non-equilibrium) line absorption coefficient and
$k_{net}\,$  is the
``net'' (continuum plus line) absorption   coefficient.

Since
$$
I_c (\nu) = B_{\nu}(T_e)\, [1 - {\rm exp}(-\tau_c)]\;, \eqno     (2.1.5)
$$
the relative line intensity at the line center is given by:
$$
r\,(\nu_0) \equiv
{{I(\nu_0) - I_c(\nu_0)} \over {I_c(\nu_0)}} =   \eta\, {{1 - {\rm
exp}(-\tau_{net})} \over {1 -
{\rm exp}(-\tau_c)}}   - 1 = \eta\, {{1 - {\rm exp}(-\tau_c - \tau_l)} \over {1
-
{\rm exp}(-\tau_c)}} - 1 \;. \eqno (2.1.6)
$$
All $\tau$'s and $\eta$ here  and following are at the central line   frequency
$\nu_0$, with the subindex omitted.
\ls
\ls
\centerline{\it {2.2. Non-Equilibrium States of a Quantum Transition}}
\ls
There is another commonly used parameter, besides $\beta_{12}$, to
describe deviations of a
quantum transition from equilibrium: the   excitation temperature   $T_x$,
which is defined by the
Boltzmann-like   formula for the population ratio of the two levels: $$
{{N_2/g_2} \over {N_1/g_1}} \equiv {{n_2} \over {n_1}} = {\rm
exp}\,(-h\nu_0/kT_x)\;, \eqno (2.2.1) $$
or
$$
T_x = {{h\nu_0/k} \over {{\rm ln}(n_1/n_2)}}\;. \eqno (2.2.2)
$$
In this subsection and in Fig.~1 we denote by
$n_i$ the population density per one degeneracy sublevel: $n_i \equiv
N_i/g_i$, with $g_i$ being the degeneracy of the level $i$; this   should not
lead to a confusion with the {\it principal quantum   number} for which the
script $n$ (without subindex) is used   throughout the paper. Subindex 1
denotes the lower and 2 the upper   level of the transition.

Since
$$
{{n_2} \over {n_1}} = {{b_2} \over {b_1}}\,{{n_2'} \over {n_1'}}\;,   $$
where primed symbols denote again the LTE values, we have the
following relation between
$\beta_{12}\;$ and $T_x$:
$$
\beta_{12} = {{1 - {\rm exp}(-h\nu_0/kT_x)} \over {1 - {\rm
exp}(-h\nu_0/kT_e)}} \;. \eqno (2.2.3)
$$
For low frequency transitions ($h\nu_0/k \ll T_e\;$), and when also
$|T_x|\,$\gapprox\thinspace $T_e$,
$$
\beta_{12} \approx {{T_e} \over {T_x}}\;. \eqno (2.2.4)
$$

The behavior of $T_x$ and $\beta_{12}$, as functions of $n_1/n_2$,     is
illustrated in Fig.~1 for
the case of $T_e = 10,000\,$K and   $h\nu_0/k = 100\,$K. When the
population ratio $n_1/n_2\,$
decreases   from $\infty\,$ (all the atoms in the lower level) to 0
(all the  atoms in the higher
level), $T_x\,$ first increases from 0 to $\infty\,$, then undergoes a jump to
$-\infty$ at $n_1 = n_2$, and then increases to 0. The $\beta_{12}\,$ parameter
decreases
monotonicly from   $\beta_{12}^{max} = 1/(1 - {\rm
exp}(-h\nu_0/kT_e))$ to $-\infty$, passing through 0 at $n_1 = n_2$.

The classification of the excitation states of a transition  is most natural
in terms of temperatures
(Strelnitski 1983), see   Fig.~1. $T_x = T_e\,$ is obviously the
state of {\it thermalization}, and
$\beta_{12} = 1\,$ in this state. When $T_e < T_x < +\infty$, we refer to the
condition as one
of {\it overheating} of the transition 1-2; it   corresponds to $0<
\beta_{12} <
1$. The case $T_x
< 0\,$   (with $\beta_{12}\,$ also being negative here) is the case   of {\it
population   inversion}.
Finally, when $T_e > T_x > 0\;$ (corresponding to   $\beta_{12} >1$),   we
call this condition {\it overcooling} of the transition 1-2.   The   choice of
this
nomenclature will become clearer as
we analyze  the observable effects of  the corresponding non-LTE   states
below (see also Strelnitski 1983).

\ls
\ls
\centerline{\it {2.3. Non-Equilibrium Populations in Hydrogen}}
\ls
Storey and Hummer (1995) have recently completed a detailed
set of calculations of the hydrogen level populations for levels $n \le 50$,
and presented the results as machine-readable files.  For Menzel's Case B
(the Lyman lines are infinitely optically thick, all other lines are thin),
calculations were made
for $2 \le {\rm log}\,N_e \le 14\,$ and for $2.7\le{\rm log\,}T_e\le 5\,$.
Figs.~2--5, plotted
with the use of the Storey and Hummer's data, show that recombining
hydrogen may acquire {\it all} the types of non-LTE  population distributions
just   described --- overcooling, overheating and inversion.

Fig.~2 presents the $b_n$ coefficients versus principal quantum   number
$n$, and as function of electron density, $N_e$, for $T_e = 10^4\,$K. The
major qualitative features of the $b_n\,$ behavior are:

(1) $b_n\,$ tends to one when $n\,$ tends to $\infty$. This     thermalization
of the highest $n\,$
levels is due to the fact that their populations are dominated by   three-body
collisional   recombinations and  collisional
ionizations, both processes involving   the same Maxwellian gas with   kinetic
temperature
$T_e$. Thus, a reservoir of quasi-thermalized levels is created at   the
highest $n$'s.

(2) $b_n$ decreases toward lower $n$, due to the ever increasing rate of
spontaneous decay which is roughly proportional to $n^{-5}\,$.
There is a  diffuse separation between those levels which are
quasi-thermalized and those which are
underpopulated and show a steep positive gradient d$b_n$/d$n$. This
boundary  progressively shifts toward lower $n$'s, as $N_e$
(and the effects of collisions) increases.

(3) At the lowest $n$'s the gradient d$b_n$/d$n$ tends to some high
positive limiting value.
Here populations are totally controlled by   radiative processes, and   the
intrinsic trends of
radiative   probabilities across the levels determine this   ``radiative
limit'' in
the slope of
$b_n(n)$ (compare this part of the $b_n(n)$ curves   with our $b_n$
calculation for the case
$N_e = 0$, shown in Fig.~2 with the broken line). The passage to the
radiative regime, at low
$n$, occurs because, at a given density, the rate of transitions   induced by
collisions with
electrons decreases with decreasing $n\,$ approximately as $n^{4\to   5}$,
whereas the rate of
spontaneous radiative   decay increases as $n^{-5}$. Thus the   relative role
of radiative
processes, {\it at a given density}, increases with decreasing $n$     very
steeply, steeper than
$n^{-9}$.

(4) At low enough densities ($N_e$\lapprox $10^5\, {\rm cm^{-3}}\,$),
d$b_n$/d$n$ {\it
changes sign} before passing to the radiative limit. As seen from
Eq.~(2.1.3), this means that
$\beta_{12}\,$   exceeds $+1$, and thus the corresponding transitions   are
{\it overcooled}. To
our knowledge, this important fact has never been   pointed out   before,
although it has been
implicitly present in the results   of earlier $(n,l)\,$ population
calculations. It
is clearly seen, for
example, in Fig.~2 of Hummer and Storey (1992). The change of sign of
d$b_n$/d$n\,$ is due to
the onset of partial ``unblurring''  of the levels' angular momentum
$l$-structure popultations at these   values
of $n$, a distribution which is blurred by proton collisions at   higher   $n$.
We interpret it as follows.
Unblurring begins when the   radiative decay rate of an $(n,l)\,$   sublevel
becomes comparable
with   the collision rate with protons. Since spontaneous decay rates
increase with decreasing
$l$, the low $l$ sublevels of a given $n$   level are unblurred the   first. At
any given density,
therefore, there is a transitional group of levels in which more and   more
$l$-sublevels (beginning
with the smaller $l$ values) are unblurred as one looks to smaller   $n$.
Blurring of the $l$-sublevels enables high $l$ atoms on a level $n$   to decay
faster, via  lower $l$ sublevels,
than the $\Delta l = \pm1\,$ selection rule normally allows. The more   a level
is unblurred the higher is the $l$ which looses
this ability to decay more rapidly.   As a result, the lower the   principal
quantum number $n$ the
more its $b_{nl}$ factors reflect higher populations in its higher   $l$
sublevels. This effect   is
seen in Fig.~3, plotted with the numerical data of Storey and   Hummer
(1995) for $T_e =
10,000\,$K and $N_e = 10^3\, {\rm cm^{-3}}$.   The effective $b_n$   factor
for a level $n$ is
derived from the statistically weighted, calculated $b_{nl}$ factors: $$
b_n = \sum_{l} {{(2l+1)} \over {n^2}}\,b_{nl}\;. \eqno (2.3.1)
$$
Since high $l\,$ sublevels have higher statistical weights $(2l+1)$,   the
increase of their
$b_{nl}\,$ factors is more significant for the   weighted value of   $b_n\,$
(2.3.1) than the decrease
of the $b_{nl}\,$   factors of low $l\,$ sublevels. As a result, {\it
unblurring
increases the
effective $b_n\,$ value of the whole $n$-level}. And   because the   degree
of unblurring increases
with decreasing $n$, a {\it negative} gradient d$b_n$/d$n$ may result   for
the partly   unblurred
group of levels, as is seen in Fig.~2.  We note that those previous
calculations
which neglected the
angular momentum state population distributions (e.g. Walmsley 1990)
show no evidence for ``overcooling.''

These four features of the $b_n$ behavior determine the shape of the
$\beta_{n,n+1}$ vs. $n$ curves for \al-lines, shown in Figs.~4{\it   a,b} for
several values of $N_e$, and the shape of the
$\beta_{n,n+1}\,$ vs. $N_e$ curves shown in Fig.~5 for several
\al-transitions.  Comparing Figs.~4{\it a,b} with Fig.~2 is instructive.

(1) At the high values of $n$ the positive slope of
$b_n$ vs. $n\,$ results in an {\it inversion} of populations ---
$\beta_{n,n+1} < 0$.

(2) In a transitional set of $n$ values $b_n$ levels   off and then changes
slope due to the unblurring of $l$ states at progressively smaller $n$
values; here  the  inversion first diminishes,   and then disappears.
If the density is low enough ($N_e$\lapprox   $10^5\, {\rm cm^{-3}}\,$), the
unblurring of $l\,$ states first reduces the inversion to overheating   ($0 <
\beta_{n,n+1}
< 1\,$), then leads to thermalization ($\beta_{n,n+1}  = 1\,$), and   finally
---
to overcooling ($\beta_{n,n+1} > 1$).

(3) The final transition of $b_n\,$ to the ``radiative limit,'' at     still
lower
$n\,$, stops the
trend towards overcooling and brings   $\beta_{n,n+1}\,$ down to the
slightly overheated values
typical for the ``radiative limit'' at low $n$.

At low densities, therefore,  the $\beta_{n,n+1}\,(n)$ vs. $n$ function   has
two extrema --- a minimum due to
inversion at high values of $n\,$, and a maximum due to overcooling   at
lower $n$'s. At densities $N_e$\gapprox $10^5\, {\rm cm^{-3}}$   there is
only the inversion minimum because ion collisions at these high densities
blur the $l$-structure  down to very low $n$'s,   and thereby remove the
effective cause of overcooling.

Fig.~5 illustrates the behavior of the $\beta_{n,n+1}\,$ factor as a   function
of $N_e\,$ for some
selected \al-transitions. Qualitatively, the  $\beta\,(N_e)\,$   functions are
similar to the
$\beta\,(n)\,$   functions: high $n$ lines (higher than $n \approx   20$) have
two extrema --- a
minimum due to inversion at high densities and a maximum due to
overcooling at low densities.  Lower $n\,$ lines
show only inversion, and this at increasingly higher densities (and
diminishing amplitude) for lines
of decreasing $n$.
\ls
\ls
\centerline{\it {2.4. Thermodynamic Considerations}}
\ls
The creation of stable conditions for inversion, overheating, or   overcooling
requires cycles of
population transfer which, from the thermodynamic   point of view,   operate
as ``heat pumps''
(Strelnitski 1983). The system   of population transfer cycles in   hydrogen is
rather complicated but   we can infer
some general principles. The two ultimate high temperature and   low
temperature reservoirs that
interact with the system of hydrogen   levels and provide free energy   to
create and maintain
non-equilibrium populations are, respectively, the ``hot'' ionizing     UV
radiation field and the
``cold'' radiation field at the lower   frequency recombination   lines. Note
that
this difference in the
high-frequency and the low-frequency radiation temperatures arises     not
because the central star
is hot: a diluted Planck spectrum with   {\it any} color temperature
possesses this gradient of
radiative   temperatures (Strelnitski 1983, Sobolev \et\ 1985). In   such a
radiation field the net
cycle of quantum transitions is always direct   (distributive) ---   high
energy
photons are split up to
lower energy   photons --- this constitutes the content of the   Rosseland
theorem   (Rosseland
1926; see also Strelnitski 1983, and Sobolev \et\ 1985). This   cycle of
fluorescence makes up the
grand radiative-radiative (RR)   pumping cycle in the recombining   hydrogen
in HII regions, and
tends to produce a global population inversion due to the steep     increase
of the sink (spontaneous
decay) rate with the decreasing   $n$. However, besides the ultimate
source and sink radiative
reservoirs, there are other, intermediate, reservoirs involved with     the
process: the recombination
transition levels themselves, with their   excitation temperatures, and the
Maxwellized gases of
electrons and   protons with their kinetic temperatures $T_e$. The heat
exchange between   these
reservoirs provides secondary cycles of population transfer.   Each   of the
two collisional
reservoirs, free electrons and protons,  have different degrees of
effectiveness in their interaction
with the different hydrogen levels   and sublevels, and because of   this
difference they modify the
pattern of the primary  (RR) cycles and can   even change, locally,   the {\it
sign} of the cycles, producing   overcooling.

We note that only {\it one} net link --- the source  link or the   sink link
--- can be collisional in these secondary cycles; the other link must be
radiative. In other words: pumping   cycles in hydrogen can be of   the
radiative-radiative (RR),
collisional-radiative (CR), or radiative-collisional (RC) type, but   they can
not
 be of the
collisional-collisional (CC) type because only {\it one} temperature,    $T_e$,
is associated with collisions in HII regions.
\ls
\centerline{3. HYDROGEN MASERS}
\ls
\centerline {\it {3.1. Maser Gain}}
\ls
The overheating and inversion of populations occur when
$\beta_{12}\,$ is smaller than its
LTE value $\beta_{12}' = 1$.  It is seen from
Eqs.~(2.1.2), (2.1.4) and (2.1.6) that $\eta\,$ is then   higher than   its LTE
value $\eta' = 1$,
$\tau_{net}\,$ is lower  than its LTE value, and in either case it increases
the relative intensity of the line.  Goldberg (1966)  demonstrated   that in
this
situation the relative
intensity of the line $r$ can increase  up to a factor of a few even   when
$|\tau_l|\,$ and $\tau_c\,$   are $< 1$.

However, to produce a  maser amplification $\gg 1$, population    inversion
$\beta_{12} < 0\,$ alone is
not sufficient.  The {\it net} optical depth $\tau_{net}\,$ must also   be
negative and its absolute value must be
considerably higher than unity.   When $\tau_{net} < 0$, both
$\eta\,$ and the numerator in Eq.~(2.1.6) are negative, and Eq.~(2.1.6)
becomes:
$$
r \approx {{|\eta|} \over {1 - {\rm exp}(-\tau_c)}}\, {\rm
exp}\,|\tau_{net}|\;.
\eqno (3.1.1)
$$
Thus the relative intensity of a   recombination line may become very   high if
the exponent
$|\tau_{net}|\,$ (the ``gain'' of the maser) is $\gg 1$.

The maser gain is determined by the length of   the amplification   path, $L$,
and the value of
$k_{net} = k_c' +   k_l'b_1\beta_{12}$ [see Eq.~(2.1.4)]. A negative   value
of $k_{net}$
can only derive from  a negative $\beta_{12}$, because all the other
quantities determining it are essentially positive. Since $k_c'$, in
particular,
is positive, free-free
absorption always reduces  the maser gain, with the amount of   reduction
depending on the frequency.

If the density of an HII region is not very high (say, \lapprox   $10^8\, {\rm
cm^{-3}}$), $\beta_{12}$ for a given transition can   attain high negative
values at the bottom of its inversion minimum   (cf.
Fig.~5), thereby increasing significantly the absolute value of $k_l   =
k_l'b_1\beta_{12}$
as compared with its LTE value $k_l'$. Yet, the density at which     maser
gain in a transition attains its maximum
does {\it not} coincide with the density of   this maximum inversion.   The
reason is that both
$\beta_{12}$ and $k_l'$ depend, and depend differently, on electron
density $N_e$, in particular
with the latter having a quadratic dependence on $N_e$.     Multiplication
of $\beta_{12}\,(N_e)$ by the steep ($\propto N_e^2$) function
$k_l'\,(N_e)$ shifts the
maximum of the product toward higher $N_e$. It also makes the range   of
densities with high
$|k_l|$ values much   narrower than the range of densities
with high $|\beta_{12}|$ values. Furthermore, since $k_l$
and $k_c'$ depend on $N_e$ differently, accounting for   free-free
absorption shifts (downward)
somewhat the value of the density where $k_{net}$ is minimum with
respect to that for $k_l$ alone.

In order to derive the optimum conditions for high-gain HRL masing    we
have performed
extensive calculations of the $k_{net}\,(N_e)$   dependence for \al-   and
\be-lines with $5 \le n
\le 100$.  Since the minimum   value of $k_{net}$ for
any line occurs at a density much higher than the density of   unblurring the
$l$-structure, our program calculating  $b_n$   coefficients
neglects the $l$-structure as unimportant.

Several illustrative cases of the calculated $k_{net}\,(N_e)$   dependence,
for   $T_e =\,$10,000
K, are shown in Figs.~6{\it a--f} (\al-lines) and 6$g$ (\be-lines).
In Fig.~7 the
$k_{net}\,(N_e)$ dependence is given for a large   interval of higher $n$. A
logarithmic scale on the ordinate has been   used to show all these lines on
one graph. Only the locus of the maximal $|k_{net}|$ values for individual
lines and the {\it envelope} of the $|k_{net}|\,(N_e)$ plots are shown for
$T_e = 5,000\,$K.
The $k_{net}\,(n)$   dependence for different densities, from
$N_e = 10^3$ to   $10^{11}\,{\rm cm^{-3}}$, and for the two temperatures is
presented   in Fig.~8.

Comparison of the $k_{net}\,(N_e)$ for the 36\al\ line (Fig.~6b) with   the
$\beta_{n,n+1}\,(N_e)$ for this line (Fig.5; the intervals   of high $|k_l|$
and
$|k_{net}|$ are indicated by the bars) shows that whereas $\beta_{n,n+1}$ is
negative over a large density   interval, with a maximum
of inversion that peaks at $N_e \sim 10^5\,{\rm   cm^{-3}}$, the   density
interval of high $|k_{net}|$ values for this line is much   narrower, and peaks
at much higher density --- $N_e \sim 10^7\, {\rm   cm^{-3}}$.

Table 1 gives $N_e^{max}$, the electron number density
at which $|k_{net}|$ attains its maximum, and the corresponding value
of $k_{net}$, for twenty \al-lines and twenty \be-lines
at $T_e =\,$10,000 and 5,000 K. The dependence $N_e^{max}\,(n)$ is shown
graphically in Fig.~9. For
the lines with $n$\lapprox 40 (these are of particular interest because
many of them have
already been observed as strong masers in \mwc), we fit the following
analytical approximation for $N_e^{max}\,(n)$:
$$
\eqalignno{
N_e^{max} &\approx 8.0\cdot 10^{15}\, n^{-5.66}\qquad(T_e = 10^4\,{\rm
K})\;, &(3.1.2)\cr
N_e^{max} &\approx 7.7\cdot 10^{15}\, n^{-5.75}\qquad(T_e =
5\cdot10^3\,{\rm K})\;.&(3.1.3)}
$$

The $k_{net}^{max}\,(n)$ dependence is approximated in the same interval
of $n$ by:
$$
\eqalignno{
k_{net}^{max} &\approx - 2.3\cdot 10^{-3}\, n^{-8.0}\qquad(T_e =
10^4\,{\rm
K})\;, &(3.1.4)\cr
k_{net}^{max} &\approx - 3.4\cdot 10^{-2}\, n^{-8.5}\qquad(T_e =
5\cdot10^3\,{\rm K})\;.&(3.1.5)}
$$

The difference between these approximations and more exact model
calcuations is under $\approx
50\%$   for $5 \le n \le 40$. The approximation (3.1.2) is drawn in   Fig.~9
as a broken line.

We point out that $N_e^{max}$ for a given transition doesn't
depend strongly on the assumed temperature.  It is seen in Table~1 and in
Eqs.~(3.1.2) and (3.1.3) which are, in fact, almost identical. For rough
estimates of $N_e^{max}$ in the whole temperature interval
5,000 -- 10,000\thinspace K their ``average,''
$$
N_e^{max} \approx 8\cdot 10^{15}\, n^{-5.7}\qquad(5\cdot 10^3\le
T_e \le 10^4\,{\rm
K})\;, \eqno (3.1.6)
$$
can therefore be used.

In contrast, $k_{net}\,(N_e^{max})$ may vary significantly with temperature,
especially
at the higher $n$ values. The temperature dependence of   $k_{net}$ is
more generally seen in Figs.~7 and 8.

The narrowness of the $k_{net}\,(N_e)$ peaks for individual lines
demonstrated in Figure~6 implies that to reach high gain masing in any
small group
of recombination lines around some $n$ an extended region at the   density
close to the ``optimum'' one for this $n$ is required.
However, if a value of
$N_e$ is fixed, rather than a value of $n$,   the corresponding maximum
value of the gain,
$|k_{net}^{max}|\,(N_e)$, is {\it not} that of the line attaining its   maximum
gain at this density: there are several lines, of smaller $n$, whose gain is
higher; likewise, there are several lines, of   higher $n$, whose gain is
smaller. This can be seen in several graphs   of Fig.~6, but especially clearly
in Figs.~7 and 8: the locus of $|k_{net}|$ values for individual \al-lines in
Fig.~7 (the two broken   lines, for two temperature values) passes {\it below}
the locus   of maximal $|k_{net}|$ values for given densities (the envelope of
the individual plots). Note that both these loci are {\it straight   lines} in
the
log\thinspace $|k_{net}|$ -- log\thinspace $N_e$ graph,   displaying that
corresponding dependencies are of the power-law   character. Fig.~8 also
shows that the lower the density, the   flatter the maximum of the
$|k_{net}|\,(n)$ function. Thus, in   contrast to the case of {\it a given
line}
where a narrow density   interval is required to attain the maximum possible
gain, in the case   of {\it a given density} (if the density is low enough,
say,
\lapprox   $10^7\,{\rm cm^{-3}}$) {\it many} lines are allowed to mase with
comparable gains.

$L_1 \equiv |k_{net}|^{-1}$ is the length
corresponding to $|\tau_{net}|\, = 1$.  Figures~6--8, Table~1 and
Eqs.~(3.1.4)--(3.1.5) thus allow us to estimate
how extended the medium of a given electron density should be along   the
line of sight to provide a maser gain $> 1$ in a corresponding HRL.
We note, however, that
all these give only an {\it upper limit} to a $|k_l|$ or a $|k_{net}|$
because the calculations were done assuming the
local line width is only due to the thermal broadening, ignoring   possible
microturbulent
broadening. If the latter is present
and described by an equivalent temperature $T_{turb}$, the value of
$|k_l|$ decreases by a factor of $[(T_e + T_{turb}) /   T_e]^{1/2}$.
The relative decrease
of $|k_{net}|$ due to any microturbulent braodening (and the corresponding
increase of $L_1$) is even
larger than that for $|k_l|$, because of the contribution from
$k_c'$ as seen in Eq.~(2.1.4). The value of $L_1\,$ may additionally   be
increased   by the
presence of large scale motions in the source (expansion,   contraction)
which diminish
the column densities of atoms at every   radial velocity (frequency), and
thereby increase the value
of   $L_1\,$ by a factor of approximately $\Delta V/\Delta v$, where  $\Delta
V\,$ is the
dispersion of radial velocities due to the large scale motions and $\Delta
v\,$ is the local radial
velocity dispersion due to the thermal motions and microturbulence.

All the data for Fig.~6, Table~1, and   Eqs.~(3.1.2) -- (3.1.5) were obtained
assuming  the media are {\it optically thin} in all recombination lines (except
for the Lyman lines which, for the Menzel's Case B, are taken to be  infinitely
optically thick; we do not consider these lines   here). Yet, since we deal
here with high-gain masers, that is with   gas which is supposed to be {\it
optically thick} in masing lines,   and since, in hydrogen, the masing and
pumping lines are intermixed and   have comparable values of absorption
coefficients, the assumption of   optical thinness in all but masing lines is
internally contradictory.  Qualitatively, the effect of radiation   trapping is
similar in the ``ordinary'' (non-inverted) and in the inverted transition
cases:
the more the radiation is trapped (the larger the modulus of the line optical
depth), the more the transitions   induced by the trapped photons compete
with spontaneous and   collision-induced transitions in controlling the
populations. The results   of population calculations for a multilevel system
with and without   account for radiative trapping may, of course, differ
drastically. It   is impossible in practice to account in a general way for
radiation trapping, not so much because it is a mathematically   complicated
nonlinear problem, but rather because the pattern of   optical depths is
peculiar to each concrete source.

Fortunately, in the important particular case of masing in a {\it   Keplerian
disk} (the case of \mwc\ and similar objects) the problem of   the {\it
pumping} radiation trapping can be finessed by supposing   that the medium
is optically thick in only one direction, that of the   chord of maximum
velocity
coherence (see Fig.~1 in our Paper~1). In   the directions perpendicular to
the disk, and also in those   directions within the disk which are far from
that
chord, the optical   depth is smaller, and it can actually be smaller than
unity,
even if   the chord optical depth is considerably larger than unity. Since
these directions occupy, all together, much larger solid angle than   the
directions close to the  chord, spontaneously radiated photons   can escape
from the medium almost freely, providing an efficient   pumping for
high-gain, optically thick masing into relatively small solid angle {\it along}
the chord.  There is one further fact providing confidence in this general
approach: the successful modeling of the masing observed in the case of \mwc\
(cf. Paper~3).

If the modulus of the optical depth along the chord of maximum
amplification is large enough, the rapid increase of maser intensity   creates
a special kind of radiation trapping problem --- maser   saturation.
\ls
\ls
\centerline {\it {3.2. Saturation}}
\ls
Saturation occurs when the growing maser radiation   begins to influence the
populations of the transitions significantly.  No general solution exists for
the
problem of   maser saturation for the same reason why   no general solution
is possible for radiation trapping in a system with positive optical depth.
Yet, as it will be seen below
and in Paper~3,   saturation may be occuring in these systems, and
understanding its role will play a crucial part in interpreting the observed
properties of masing HRLs, and in predicting the observability of   other
lines. In this section we consider two basic questions   concerning saturation
in HRLs: (1) how saturation in a masing line  affects the level populations
(thus the gain) of the masing   transition itself; and (2) how, in a multi-line
HRL maser, saturation   in one line affects the populations and the maser
gains in adjacent   lines. We also derive a criterion for suspecting when
saturation has occured in a masing HRL.

The first question has been intensively studied in the context of   molecular
masers. The
condition of saturation is that the maser   radiation density influences
``noticeably'' the
population difference $\Delta N_{21}$ between the upper (2) and lower   (1)
maser levels.
Ideally one needs an analytical expression   for $\Delta N_{21}$ as a
function of all the basic
processes   influencing  the populations. In practice this goal is
unattainable
when the number of
interacting levels is $>3-4$.  Several simplifying   approaches have
therefore been proposed to get around this problem   (see the   discussion in
Strelnitski, 1993).  In all these approaches   the two {\it relaxation}
processes,  radiative (maser) and    collisional   transitions between the
maser
levels themselves,  are   accounted for almost exactly. It is much   more
difficult,   however, to
provide for a complex system a simple yet adequate   approximation   for all
the {\it pumping} processes.

In one approach, pumping is represented by four phenomenological
coefficients for the upper
and lower maser levels: the ``income''   coefficients $\lambda_2\,$   and
$\lambda_1\,$
(cm$^{-3}\,$s$^{-1}$)   and the ``outlay'' coefficients $\Gamma_2\,$   and
$\Gamma_1\,$  (s$^{-1}$).  Omitting subindex ``12'' where it cannot lead to
misunderstanding,   and taking for simplicity the
degeneracies of the maser levels to be equal, the solution of the   statistical
equilibrium equations in
this ``$\lambda, \Gamma$''-approach can be represented as follows
(e.g.: Strelnitski, 1993):
$$
\Delta N = {{\Delta N_0} \over {1 + {{BJ} \over {\Gamma +     C_{21}}}}} =
{{\Delta N_0}
\over {1 + {{J} \over {J_s}}}}\,, \eqno   (3.2.1)
$$
where
$$
\Delta N_0 = {{{{\lambda_2} \over {1 + \Gamma_2/\Gamma_1}} -
{{\lambda_1} \over {1 + \Gamma_2}/{\Gamma_1}} \over {\Gamma +
C_{21}}}} \eqno
(3.2.2)
$$
is the unsaturated population difference; $\Gamma \equiv
(\Gamma_2^{-1} +
\Gamma_1^{-1})^{-1}$ is the harmonic mean   value of the decay rate   of
the two levels; $B$ is the Einstein   coefficient for induced   radiative
transitions between the maser   levels  ($B \equiv B_{12} = B_{21}$ since
we assumed the   degeneracies of the two   levels to be equal); $J$ is
the actual radiation intensity averaged over the profile of the   absorption
coefficient and over directions;
$$
J_s \equiv (\Gamma + C_{21})/B \eqno (3.2.3)
$$
is the so called ``saturation intensity;'' and $C_{21}\,$  is the rate of
collisional relaxation between
the maser levels.  To obtain   this form of eqs.~(3.2.1) and (3.2.2)   we
ignored the slight
differences   between $C_{12}\,$ and $C_{21}$. Strictly speaking,    then,
these equations are only
valid for transitions with $h\nu \ll   kT_e$ which, at $T_e \sim   10^4\,$K,
comprise all hydrogen
\al-lines down to $\approx\,$5\al.

Eq.~(3.2.1) shows that maser radiation begins to influence the maser
level  populations when the
rate of its interaction with the maser levels, $BJ$, becomes  comparable
to the higher of the two rates: the   rate of collisional   relaxation between
the
levels, $C_{21}$, and the rate of
the decay of the populations from the maser levels in the pumping
processes, $\Gamma$. (It is worth noting that
harmonic mean is always  smaller than the smallest of the two   averaged
numbers, and equal to
one half of each when they are equal.) The {\it net} effect   of the   maser
radiation is the
transfer of population from the upper   to the lower state. It is   seen in
Eq.~(3.2.1) that in the limit
$J   \gg J_s\,$ the population difference $\Delta N \to 0$, and thus
$(N_2/N_1) \to 1$. Note,
however, that in this limit,   $\Delta N \propto J^{-1}\,$ [see
Eq.~(3.2.1) with $J \gg J_s$]. This dependence of $\Delta N\,$ on     $J\,$
reduces the maser
amplification from an exponential [Eq.~(3.1.1)]   to a linear   dependence.
The effect is easily
understood by noting that the radiative   transfer equation for a
sufficiently
strong maser (where the
spontaneous  term is negligible when compared to the induced term)
reduces to:
$$
{{dI} \over {dz}} = {{h\nu} \over {4\pi}}\,B\, \Delta N\,I\,.
$$
When $\Delta N$ is independent of the intensity $I$ (unsaturated   regime),
integration gives
an exponential growth of $I$ with $z$;   when $\Delta N \propto   I^{-1}\,$
(saturation), $I$
depends on $z$   linearly (see also general reviews of maser theory,   e.g.,
Strelnitski, 1974; Reid and Moran, 1981; Elitzur, 1992).

The major simplifying assumption of the
``$\lambda,
\Gamma$''-approach is that $\lambda_2\,$ and $\lambda_1$ do not
depend on the maser level
populations. This leads eventually to the expression (3.2.1) which states
that $(\Delta N/\Delta
N_0)$ does   not depend on $\lambda$'s but only on $\Gamma$'s and $C$
--- a   radical
simplification. The fewer the number of the levels involved   with   the
pumping, the worse the
approximation, because in such a  few-level system the  non-maser   level
populations --- the
source of ``income'' for the maser levels --- depend sensitively on   the
populations of the maser levels.

In the case of HRLs pumped by recombinations, cascading and     collisions,
every level is fed by
{\it many} other levels. If there   is only one masing transition in   the
system,
the change of its
populations caused by saturation should not produce any serious
distortion of the reservoir of
pumping. Moreover, even if masing and   saturation occur in {\it   several}
adjacent lines in the
same region   of space, this will hardly change the reservoir of   pumping for
individual transitions
because saturation only re-distributes   populations between two   adjacent
levels, without bringing
populations   in or out of the reservoir as a whole. Thus the   ``$\lambda,
\Gamma$''-approximation should give an adequate   qualitative   description
of the $\Delta N (J)$
dependence for individual   masing recombination lines, in spite of   the fact
that in hydrogen the
reservoir of populations for pumping may heavily overlap with masing
transitions.

The hydrogen levels we consider here decay primarily through
spontaneous and collision-induced transitions to other levels;  thus,
$$
\Gamma \approx A_t + C_t\,,\eqno (3.2.4)
$$
where $A_t \equiv (A_{t1}^{-1} + A_{t2}^{-1})^{-1}\,$ and $C_t \equiv
(C_{t1}^{-1} +
C_{t2}^{-1})^{-1}\,$ are the harmonic mean values of   the total   Einstein
$A$ coefficient and
of the total collision rate   for the two maser levels respectively.   Using
Eq.~(3.2.4) and the
relation between   the Einstein $A$ and $B$ coefficients, we can   rewrite
Eq.~(3.2.3) as
$$
J_s \approx {{2h\nu_0^3} \over {c^2}}\,{{A_t + C_t + C_{21}} \over
{A_{21}}}\,, \eqno
(3.2.5)
$$
where $A_{21}$ is the Einstein coefficient for the maser transition,
$\nu_0\,$ is the
transition frequency and the constants have their usual meanings.

To estimate the degree of saturation in an observed masing line, we    have
to compare an estimate
of $J$ in the source with $J_s$ given   by Eq.~(3.2.5). A key and   complex
issue is that of the
radiation's solid angle, but we are able to sidestep the matter    using the
following argument (cf.,
Strelnitski, 1984). On the one hand, unless the geometry of the     source is
very special, the solid
angle of the maser beam should be close to
$$
\Omega \approx {{\sigma}\over {l^2}}\,,\eqno (3.2.6)
$$
where $\sigma$ is the observed emitting surface area, and $l$ is     the
amplification path
length. On the other hand, the solid angle   under which we see the   source
is $$
\omega \approx {{\sigma}\over {D^2}}\,,\eqno (3.2.7)
$$
where $D$ is the distance to the source.

Thus, $\sigma$  cancels out in the equation connecting $J$ with   the
observed flux density $S$:
$$
J = I {{\Omega} \over {4\pi}} = {{S} \over {\omega}}\,{{\Omega} \over
{4\pi}}   = S {{D^2}
\over {4\pi l^2}}\;, \eqno (3.2.8)
$$
where $I$ is the specific radiation intensity averaged over the     profile of
the absorption
coefficient. From Eqs.~(3.2.5) and (3.2.8)  we obtain the degree of   maser
saturation:
$$
{{J} \over {J_s}} \approx {{c^2\,D^2} \over {8\pi h \nu^3_0}}\,     {{A_{21}}
\over {A_t + C_t
+ C_{21}}}\, {{S} \over {l^2}}\;. \eqno   (3.2.9)
$$

Eq.~(3.2.9) gives an estimate of the degree of saturation in terms of     the
transition constants
$\nu_0,\, A_{21},\, A_t$, the model   parameters $C_t,\, C_{21}$
(determined by the model
density of the   emitting region), the observed flux parameter $S$, and   the
``semi-observed'' parameter $l$ which, though not directly   observable, can
in principle be reliably
estimated from direct interferometry of the source structure in   the plane of
the sky   and the geometrical model of the source.
We emphasize that transition to saturation is a {\it gradual}     process. It
is
seen from Eq.~(3.2.1)
that to pass from an   unsaturated regime $(J/J_s \ll 1)\,$ to a   heavily
saturated one $(J/J_s \gg
1)$, the intensity needs to increase by about two orders of     magnitude.
Thus the saturation
criterion  $J > J_s$ is no more   than an {\it order of magnitude}   criterion.

There is one case for which we can simplify
Eq.~(3.2.9)   without losing
its exactness: the case where maser radiation is   generated at the
``optimum'' density, i.e., the
density that provides the maximum (unsaturated) value of $|k_{net}|$   for
the line in
question. As  was discussed in Section 3.1, the   optimum density for
maser
amplification is much higher than the   density of maximum inversion,   and,
as Figs.~5 and 6 show,
this   optimum density is actually rather close to the density of
thermalization. At such high
densities, $A_t$ in Eq.~(3.2.9) can   always be ignored as compared   with
$(C_t + C_{21})$. For $5 < n < 35$ and $T_e = 10^4\,$K, we use   the
following approximation:
$$
\delta \equiv (C_t + C_{21}) \approx 3\cdot 10^{-9}\,n^5\,N_e\;,     \eqno
(3.2.10)
$$
which is a simplified version based on Gee's \et\ (1976) analytical
approximations for collision rates and which is exact to $\pm 5\%$.
Substituting for $N_e$  the approximation (3.1.6) for $N_e^{max}$,   and
taking $A_{21}
\approx 6.3\cdot 10^9\, n_1^{-5}\,$(s$^{-1})$ and   $\nu_0 \approx   6.6\cdot
10^{15}\,n_1^{-3}\,$(s$^{-1})$ --- common approximations valid for   \al-lines
with $n \gg 1$ ---
and finally  neglecting   $A_t$ in Eq.~(3.2.9) as compared with $(C_t   +
C_{21})$, the equation reduces to
$$
{{J} \over {J_s}} \approx 0.2\, \biggl ({{D} \over {{\rm   kpc}}}\biggr   )^2\,
\biggl
({{n} \over {10}} \biggr)^{4.7}  \biggl ({{S}   \over {{\rm   Jy}}}\biggr
)\,\biggl
({{10^{13}\,{\rm
cm}} \over   {l}}\biggr )^2\;. \eqno (3.2.11)
$$

The previous assumption, that the observed masing lines arise in regions   of
``optimum'' density, is of
course an assumption which can be wrong. One reason may be the
interaction of the
adjacent masing transitions due to saturation.   Suppose some   line   $2 \to
1\,$ mases at its
optimum density and reaches strong   saturation. We can ask how such
saturation might affect the
trends of masing in the next transition  higher, $3 \to
2$.  If the $2 \to 1$ line were not saturated, the $3 \to
2$ line would be  if not thermalized, at this high density,  at least
amplifying less strongly than its
own  optimum (lower) density would predict  (see Fig.5 for an   illustration).
According to
Eq.~(3.2.1), however, saturation of the $2 \to 1$ transition means    that
$\Delta N_{21}$ decreases
due to the transfer of some population from level 2 to level 1, and     implies
that the population of
the lower level of the  $3 \to 2$  transition decreases.   In other   words,
{\it
saturation of   the
transition $2 \to 1$ should increase the inversion of the     transition} $3
\to
2$. In thermodynamic
language, saturation of the   transition $2 \to 1$ opens a new   channel of
sink to the maser
on the transition $3 \to 2$, thereby increasing the efficiency of     its
pumping. As a result, the $3
\to 2$ line will be amplified to a   higher intensity. If it in turn
saturates, it will
increase in the
same   manner the inversion of the next higher transition, etc. The
efficiency of this upward
diffusion of the reenforced pump sink   decreases, however, at every   step,
until  eventually, at some
higher   value of $n$, the steeply increasing relative rate of   thermalizing
collisions will remove the population inversion.

Thus, a saturated high frequency masing line can ``attract'' an     adjacent
group of lower frequency
lines to mase in a region of   density higher than their unsaturated
``optimum'' densities. We will
apply   this mechanism, in Paper ~3, to interpreting some observed
properties of the multi-line
hydrogen maser in \mwc. In any particular case, however, this   qualitative
idea requires a more
exact numerical check --- a   calculation of level populations   including
induced   radiative
transitions in the lines which are supposed to be saturated masers.
\ls
\ls
\centerline{4. HYDROGEN ``DASARS''}
\ls
The effect of overcooling, demonstrated for hydrogen Rydberg levels     in
Fig.~4a and 5 and
discussed in Section 2.3, is actually a well known   phenomenon for   radio
lines of some interstellar
molecules (H$_2$CO, OH). Observationally it appears as an enhancement
of a   line seen in
absorption. The observational effect of overcooling was called   ``DASAR''
(``Darkness Amplification   by the Stimulated Absorption of   Radiation'') by
C.~Townes and A.~Shawlow.  Overcooling is sometimes   called
``anti-inversion,'' and  the
corresponding enhanced absorption --- ``anti-maser.'' The analysis of
non-LTE states of a
quantum   transition given in Section 2.2 shows, however, that   overcooling
is   really more an
analogue of overheating than of inversion (see more in Strelnitski,   1983).

We first show that for an homogeneous, ionized cloud with no external
continuum background, the observation of a line in absorption, i.e., $r < 0$,
{\it requires} $\beta_{12} > 1$ -- overcooling of the corresponding transition.

{}From Eq.~(2.1.6), $r < 0\,$ corresponds to
$$
\eta < {{1 - {\rm exp}(-\tau_c)} \over {1 - {\rm exp}(-\tau_c -
\tau_l)}}\;.
\eqno (4.1.1)
$$
It is easily seen that the right-hand side of Eq.~(4.1.1) is always     $< 1$,
whatever the value (or
sign) of $\tau_l$, and hence $\eta < 1$, which is only possible   according to
Eq.~(2.1.2) if
$\beta_{12} > 1$.

Thus, in the framework of this simple model, the very observation of a
recombination line in
absorption on the proper free-free   continuum of the cloud is a   direct
indication that the transition
is   overcooled. More complicated circumstances, however --- a hotter
external   continuum
background for example, or some peculiar morphology of the cloud, like
perhaps if the   kinetic
temperature decreases radially outward --- can produce an absorption
recombination line without
overcooling.  Usually such cases can be identified by a detailed   analysis of
the source.

Unfortunately, the reverse statement, namely that $\beta_{12} > 1$
(overcooling) {\it always}
results in $r < 0\,$ (a line seen in   absorption), is not true. To   see when
overcooling does produce
an   absorption line in our model, we consider three particular   cases:

(1) $\tau_c \ll 1,\; \tau_l \ll 1$;

(2) $\tau_c \ll 1,\; \tau_l$\gapprox 1;

(3) $\tau_c$\gapprox 1.

In the first case, expanding the exponents in Eq.~(2.1.6) to first   order
terms
and using Eqs.~(2.1.2) and (2.1.4), we have:
$$
r \approx \eta {{\tau_{net}} \over {\tau_c}} - 1 = {{k_l'} \over
{k_c'}}\,b_2 \;.
\eqno (4.1.2)
$$
Since all the terms on the right-hand side are positive, $r$ can only   be
positive in this case. In
other words, only {\it emission}   recombination lines can be   observed from
an optically thin cloud with no external continuum   background.

Case (2) can occur for high frequency recombination  lines   ($n$\lapprox
30), for which $k_c' \ll
k_l'$   (see Fig.10). Ignoring $k_c'$ in Eq.~(2.1.2) and expanding the
denominator on the
right-hand side of Eq.~(2.1.6) to first order, we have, as the   condition of
$r
< 0$ for this case:
$$
\tau_c > {{1 - {\rm exp}\,(- \tau_l)} \over {\beta_{12}}} \;. \eqno     (4.1.3)
$$
Thus, for high frequency lines overcooling can produce an absorption line
even in a cloud
optically thin in the continuum, however, the line optical depth of the cloud
needs to be \gapprox 1
and its continuum optical depth should satisfy the condition above   (4.1.3).

Finally, we consider case (3), first in its asymptotic form, when   $\tau_c \gg
1$. Eq.~(2.1.6) reduces then to $r \approx \eta - 1$, and   the condition $r <
0$ becomes $\eta < 1$
which, according to   Eq.~(2.1.2), is identically satisfied if   $\beta_{12} >
1$.
Thus, for   a
homogeneous cloud optically thick in continuum an overcooled
recombination line will always
be observed in absorption (whatever  the value of  $\tau_l$).

Now we show that this is true even for a {\it moderately}   optically   thick
cloud ($\tau_c$\gapprox 1). In this case both numerator and   denominator
in Eq.~(2.1.6) are  less than 1 and greater than $(1 -
{\rm e}^{-1}) \approx 0.63$. Thus the condition $r < 0$  in   this   case
becomes $\eta < 0.63$.
After some algebra on equation  (2.1.2) for $\eta$, and remembering   that
$b_2/b_1 < 1$ when
$\beta > 1$, the $r < 0$ condition becomes:
$$
{{k_l'b_1} \over {k_c'}}\; {_>\atop{^\sim}}\; {{0.6} \over     {\beta_{12} -
1.6}} \;.
\eqno
(4.1.4)
$$
Taking $\beta_{12}$ (at its maximum) from Fig.~5, $b_1$ from     Fig.~2,
and $k_l', k_c'$ from
Fig.~10, it is easy to see that   Eq.~(4.1.4) is satisfied for all the
transitions
which can be
overcooled in principle, $n$\gapprox 20 up to at least $n = 56$, seen   in
Fig.~5. Thus,
for all these transitions   (and in fact for higher $n$ transitions   as well)
overcooled
lines   will be observed in absorption, if $\tau_c$\gapprox 1.

Observation of an absorption line on the proper free-free   continuum   of an
HII region is the most
direct way to reveal overcooling, but it is not the only  possible   way. At
low
optical depths
(Case~1) there is another obvious way: to observe {\it more than one} line
in the interval of
$n$ where overcooling is theoretically   predicted and to determine   the
decrement of $r$ with
respect to $n$. According to Eq.~(4.1.2), in the optically thin case   $r
\propto   b_n$, hence, the
observed {\it sign} of the decrement of $r$ should show whether or   not
overcooling takes place.
\ls
\ls
\centerline{5. PROSPECTS}
\ls
In this Section we address the observational prospects for HRL   masers,
lasers, and dasars. Our analysis of masing shows that there   are at least
two preliminary questions to be addressed when   considering an object, or
a class of objects, as a potential candidate: (1) what are the {\it densities}
of ionized hydrogen in the source?  and (2) what is the linear scale, or,
more exactly, the column density   per
thermal line width, at each density? The answer to the first question shows us
which lines might be   inverted or
overcooled   in the
source, the answer to the second --- whether or not these non-LTE   lines
can   become high gain masers or
detectable dasars.

We examine now, with these two test questions, several classes of   objects.
This list of candidates should not be regarded as a complete   one: we only
give several examples. We do not discuss here the   objects similar to \mwc\
--- hot emission line stars with neutral   disks ionized on their surface. Some
observational prospects   regarding this class of sources will be presented in
Paper~3.
\ls
\ls
\centerline{\it 5.1. Compact HII Regions}
\ls
Simple estimates show that among the compact HII regions associated
with the sites of massive
star formation, only the densest , ``ultracompact,'' ones can have   optical
depths in HRL close or surpassing unity by absolute value as   required  for
masing. Typical electron densities in ultracompact HII   regions
are $N_e \sim 10^4$--$10^5\,$cm$^{-3}$, but in some of them an {\it
average} density is estimated to be as much as $N_e \approx 3\cdot
10^5\,$cm$^{-3}$ (Wood   and Churchwell 1989; Kurtz \et\ 1993).
Figure~8 shows that the lines in a broad interval, from $\approx
35\alpha$ to $\approx 60\alpha$, have comparably high $|k_{net}|$ at   this
density, which may reach $\sim 10^{17}\,{\rm cm^{-1}}$, if the   temperature
is somewhat lower than 10,000~K. Since many of these   ultracompact HII
regions have dimensions $L$\gapprox $10^{17}\,$cm,   the maser gain
$|k_{net}|\,L$ may surpass unity in these lines, even   if the velocity
dispersion reduces the effective value of $|k_{net}|$   by a factor of 1.5--2.
Thus, some increase of intensity and line   narrowing due to masing is
anticipated in this  interval of   \al-lines from the most compact HII regions.

Some of these regions are highly inhomogeneous. Since chances of
high-gain masing increase, in general, with the increasing density,   masing
in lower $n$ lines is also quite possible from local density   enhancements,
even if such zones of higher density are considerably   smaller than the
dimensions corresponding to the average density of   the object.

HII regions associated with evolved objects, for example planetary nebulae,
have in general lower densities.  Nevertheless studies of forbidden line
transitions (e.g., Osterbrock, 1989) show that in some cases like IC4997
there are regions with densities up to $10^6\,{\rm cm^{-3}}$.

Effects of masing can be studied by comparing the
intensities and widths of several \al-lines over a broad range of $n$, and
also by comparison with the \be-lines from nearby $n$ (see Paper~3). The
analysis would help
specify more exactly the physical parameters of the HII region.
\ls
\ls
\centerline{\it 5.2. Wolf-Rayet Stars}
\ls
The idea to search for HRL line masers in Wolf-Rayet (WR) stars
originated with
H.E.~Matthews (1992; private communication).   These stars   possess
strong winds and
expanding ionized circumstellar disks which might promote population
inversions and maser
amplification similar to those in the disk of \mwc. However, the     first
attempts to find mm and
submm hydrogen masers in WR   stars gave negative results (Matthews
1994; private communication).

The most probable reason is insufficient optical depth in the lines   (30\al\ -
26\al) studied. They
require $N_e \sim 10^7$--$10^8\; {\rm   cm^{-3}}\,$ (see   Table~1) for
effective masing --- but
these densities are not sustained in large enough {\it   column}   densities in
a WR disk outflow.
If we approximate the   ionized envelope of a WR star as a   spherically
symmetric,
\gapprox $1000\,$\kms\ constant velocity (thus $N_e \propto R^{-2}$)
outflow  with a mass
loss rate of
\lapprox $10^{-4}\,{\rm M}_{\odot}\,{\rm yr^{-1}}$, it is easy to   show that
the
linear scale   of
the region of density $N_e$ will be $R(N_e)$\lapprox $2\cdot
10^{18}N_e^{-1/2}\,$cm. At
the optimum density of the H26\al\ line, for example,  where   $N_e^{max}
\approx 1\cdot 10^8\;
{\rm cm^{-3}}$, we find $R(N_e^{max})$ \lapprox\  $2\cdot   10^{14}\,$cm.
Table~1 gives
$k_{net}^{max} \sim - 10^{-14}\; {\rm cm^{-1}}$ for this line, but,     in
calculating the optical
thickness we must reduce this value by   the ratio of the radial   velocity
dispersion in a WR
envelope, $\Delta V \sim 10^3\;$\kms, to the thermal line width,
$v_{th}\sim 20\,$\kms\   (see
Sect.~3.1). As a result, the net optical depth in the line   reduces   to
$\tau_{net} \approx
k_{net}\,R(N_e^{max})\,(v_{th}/\Delta   V) \sim - 10^{-2}$. In the   same way
we can show that
$|\tau_{net}|\ll 1\,$ for other mm and submm lines. Though     $|\tau_{net}|$
does increase  with decreasing $n$, it becomes   \gapprox 1 only at
$n$\lapprox 10. Thus, high gain lasers in hydrogen recombination   \al-lines
from the   expanding shells of WR stars
are most promising for the IR lines   with $n$\lapprox 10.

At this high a frequency a different source of line emission can   be   even
more effective at
producing laser lines in WR stars:  the  {\it atmosphere}.  In the   standard
model of a WR  star the
non-expanding atmosphere  consists of a turbulent layer where $N_e \sim
10^{11}$--$10^{13}\,
{\rm cm^{-3}}$ and which is $\sim 5\cdot   10^{11}\,$cm thick. Such
densities are optimum for lasing in
$n$\lapprox 7 lines (see Table~1). Multiplying the width of this   layer   by
our
calculated
values of $k_{net}^{max}$, we find very large optical depths; with   $Ne \sim
10^{12}\; {\rm
cm^{-3}}$, for   example, the optical depth of   B\al\ should be   $\sim -
200$!  Even assuming the
lower limit to the density in this layer (deduced by Aller from   optical
spectroscopy) of  $N_e \sim
10^{11}\; {\rm cm^{-3}}$,  we still have optical depths of $-0.2$,   $-3$, and
$-2$ for the B\al, H6\al,
and H7\al\ lines respectively, with B\al\ being the most promising   line if
the
density is higher than this lower limit.

The observability of an HRL laser is, however, strongly contingent   upon its
geometry. If the solid angle of lasing is not small enough,   this emission
tends to be lost on the stronger, spontaneous emission   background. This
problem is discussed in more detail in Paper~3.
\ls
\ls
\centerline{\it 5.3. Be Stars}
\ls
Smith et al. (1979) suggested that Be stars might be likely   candidates for
lasing in the infrared
hydrogen lines, although at that time the lack of theoretical high   density
population data precluded
a quantitative discussion. These complex objects have ionized   Keplerian
disks (Slettebak \et\ 1992),  high mass loss rates,   photometric variability,
and, most
important for our considerations, inferred densities in the region   emitting
Balmer lines of $N_e \sim 10^{11}$--$10^{13}\, {\rm   cm^{-3}}$, over
dimensions $\sim 10^{12}$--$10^{13}\,$cm. These   parameters
are similar to those   in the turbulent layer of a WR atmosphere,    and
similar considerations hold
here too, suggesting they are promising candidates for lasing in the   near IR
 recombination
\al-lines ($n$\lapprox 10).
\ls
\ls
\centerline{\it 5.4. Active Galactic Nuclei and Starburst Galaxies}
\ls
Enormous masses of ionized hydrogen are present in the central parts    of
galaxies with active
nuclei (AGN), while starburst galaxies, which may sometimes be the result
of   two galaxies in collision, often have bright compact HII regions outside
the nucleus.  In the case of AGN's we consider the possibilities of   maser
action in two types of regions (e.g., Osterbrock 1989). In the   so-called
"Broad   Line Regions" (BLR)
$N_e \sim 10^8$--$10^{10}\, {\rm cm^{-3}}$, and may range between $L
\sim   0.01$--$0.1\, {\rm
pc}$ (Seifert 1 galaxies) to $\sim 1\, {\rm pc}$ (quasars).  In the   "Narrow
Line Regions"
(NLR) $N_e \sim 10^2$--$10^4 \, {\rm cm^{-3}},\, L\,$\lapprox 1000   pc.
However, the densities above do not apply to these linear scales   but ruther
to much smaller ``clumps'' or cloudlets--- the filling   factor should be much
less than one.

At the densities typical for BLR, H\al-lines with $n \approx 10 - 25$   have
maximum local
maser gain (Fig.~8). Taking   the values of $k_{net}^{max}$ for these   lines
from Fig.~6 or Table~1, we
see that linear   scales of only $\sim 10\,$AU (25\al) to $\sim   0.01\,$AU
(10\al) are   needed to
make these lines optically thick, if line broadening is thermal. If   the
individual cloudlets in BLR are
of these dimensions or bigger, then laser emission in these submm and
IR lines is anticipated, and would be seen as relatively strong   narrow
features centered at the radial
velocities of individual cloudlets. It seems quite possible. For   example, in
the Scoville and Norman (1988) model of the broad   emission line clouds as
photoionized mass-loss envelopes of giant   stars, the Str\"omgrem sphere
reaches down to a radius of $\sim   10^{14}\,$cm in the stellar envelope.
With the density there being   $\sim 10^9\,$cm$^{-3}$, the 10\al-15\al\ lines
should have high   unsaturated laser gains --- on the order of tens.

The NLR gas seems to be a less likely prospect for high-gain masing   in
HRLs. The
densities of the NLR gas are optimum for masing in high $n$    \al-lines ---
$n$\gapprox 50. Densities $\sim 10^4\,{\rm cm^{-3}}$   have linear scales up
to $\sim 1\,$pc in AGNs (viz., the ionized   accretion disk ``halo'' postulated
by Scoville and Norman, 1995).   According to Fig.~8, the 50\al\ -- 80\al\
lines have the highest   $|k_{net}|$ at this density, but it is easy to see
that
maser gains of only \lapprox 0.1 are achievable. The regions with $N_e
\sim   10^2\,{\rm cm^{-3}}$ can be much larger, up to $\sim 1\,$kpc, but
considering the velocity dispersion in such a large region, we have   to
reduce the optical depth at  each frequency by a factor of $\Delta V/v_{th}
\approx 50$, and   $\tau_{net}$ turns then out to be $< 1$ in this case too.

Maser effects in mm HRLs are anticipated in the starburst galaxies   whose
archetype, M82, was in fact the first external galaxy where a   radio
recombination line (H166\al) had been detected (Shaver \et\   1977).
Kronberg \et\ (1985) postulate dense ($N_e \sim 10^7\, {\rm   cm^{-3}}$)
ionized shells surrounding supernovae which had exploded   inside dense
molecular clouds of the  M82-like starburst galaxies.   According to Fig.~8,
\al-lines with $n \approx 20$--35 have the   highest maser gains at this
density. With the typical linear scale   for the shells being $\sim 3\cdot
10^{16}\,$cm, Fig.~8 shows that the   gain can surpass unity, if large-scale
motions do not broaden the HRL   too much, as compared with the thermal
width, and if $T_e$ is not   much higher than 10,000\thinspace K. Though
the both latter   assumptions can be wrong, some indications of narrow,
presumably   weakly masing components in the 36\al\ line from M82 has
recently   been detected by Smith \et\ (1995).
\ls
\ls
\centerline{\it 5.4. Hydrogen Dasars}
\ls
{}From an observational point of view the major difficulty with the
overcooling phenomenon in
HRLs, as compared with inversion, is the low upper limit for $N_e$:
overcooling disappears at
$N_e$\gapprox $10^5\, {\rm cm^{-3}}$ (see Sect.~4).  At this limiting
density hydrogen \al-lines
with $n \approx 20$--$25$ are overcooled, and the Case~2 (cf., Sect.~4)
applies.  Since $\beta_{12}$ is only slightly greater than 1 for   these
lines,
Eq.~(4.1.3) shows that
$\tau_l\,$ must be $> 1$   to observe an absorption line on the   proper
continuum of an HII
region. The typical value of $k_l$ for these lines, at $N_e \sim     10^5\,
{\rm
cm^{-3}}$, is $k_l
\sim 10^{-20} {\rm cm^{-1}}$   (Fig.~10), and thus very large HII   region
dimensions, of $>
300\,$pc, are needed for   $\tau_l > 1$. HII regions with such   parameters
are unknown.

At lower densities, which might make it easier to find larger   regions, the
maximum of overcooling
shifts to the higher   $n$ lines. The values of $b_n$, $\beta_{12}$   and the
oscillator   strengths
of the lines all increase, but the $N_e^2$ factor   decreases faster,   so the
required linear scale
increases. Of known objects, we believe only AGN's can be considered
today as possible
candidates for {\it direct} observations of hydrogen dasars. If the     regions
of intermediate
density, $N_e \sim 10^4\, {\rm cm^{-3}}$,   have  dimensions   comparable
with NLR's (up to
$\sim 1000\,$ pc),   \al-lines from $n \approx 30$--$35\,$ may   acquire
$\tau_l \sim 1\,$   and appear as dasars.

The most realistic way to detect the effect of overcooling in   hydrogen,
however,  is to observe
more than one line in the $n$   range where overcooling is predicted   at a
given density, and to
determine the decrement of $r$ with $n$, as described in   Sect.~4.   This
way of inferring overcooling is applicable to any   galactic HII   region if
the
observations are done in the proper
frequency   range.
\ls
\ls
\centerline{6. CONCLUDING REMARKS}
\ls
Hydrogen, the simplest and the most abundant atom in the Universe, is
also the first known {\it
atomic} astrophysical maser. Perhaps the most striking feature of   this
maser, as
compared with {\it molecular} astrophysical masers,  is its obstinate
uniqueness:  six   years after
its discovery in \mwc\ there is still only one known strong hydrogen   maser
source, despite intense
attempts to find another. In contrast, practically every newly   discovered
molecular   maser
line was quickly detected in a multitude of similar   sources.

Any useful theory of the hydrogen maser phenomenon must   directly
address this striking fact.

In this paper we have attempted to show that, although hydrogen may   be
common,  the physical
conditions necessary for   creating a high gain HRL line maser are rather
special: for every
line (or narrow group of lines) only a limited range of density is   able to
producing the high
negative values of the local absorption   coefficients required.  Furthermore
the region with this special
density must be  adequately extended, and homogeneous enough in its
radial   velocity, to create an optical
depth $< -1$.   The velocity dispersion parameter is   an especially
important one. One of the advantages
of molecular masers is   that they are located in relatively cold,   neutral
gas
clouds and are
produced by relatively heavy particles, so the width of the local     thermal
velocity profile is, in
general, only  $\sim 1\,$\kms.   Together with the high density of   these cold
clouds, such a small
velocity dispersion  can produce over even a short distance a high   negative
 optical depth at the
line center, and thus very strong masing.  A HRL  maser, on the other
hand,  is generated in a
hot ionized gas and is radiated by the lightest particle, whose thermal
velocity dispersion is
therefore   very high, $\sim 20\,$\kms. As a result the local absorption
coefficient is relatively low,
regardless of the degree of   inversion, and so long distances are required
to create $\tau < -1$. But longer distances bring an additional velocity
dispersion to the   picture
due to  their large scale
motions --- expansion, contraction, and the like, which broaden the
frequency profile of the amplified
radiation even more,  thus hindering strong masing.

In such situations {\it chance} may become a decisive factor for creating
an {\it observable}
maser. In the case of \mwc, there are at least two lucky coincidences: (1)
a very high UV-
luminous a star, perhaps in a particular outflow phase of its life,
surrounded
by a dense, massive
neutral disk, resulting in an unusually dense, extended HII region on the
disk's surface; and (2), a
fortuitous projection of the disk to our line-of-sight as edge-on.
A Keplerian
velocity field seen   edge-on is a better ``organizer'' of the atoms' radial
velocities
than an outflow --- that is why we observe stronger mm and submm masers
from the disk of \mwc\
than from its outflow (cf. Paper~3).

Whatever the particulars of HRL line masers and lasers, nature is
fabulously inventive. Conditions favorable  to the creation of rather strong
hydrogen   masers and lasers,   or even
abnormally strong absorbers (dasars), might occur in several     classes of
astrophysical objects,
both galactic and extra-galactic. The strong hydrogen maser  in \mwc\   was
discovered
accidentally, during   a routine search for more recombination lines   in this
well-known   emission
line star,  at a time when the paradigm of ``decrease of stimulated
emission   effects toward higher
frequencies'' was strongly held.  Now,   thanks to the example of   \mwc\
coupled with a thorough
analysis of  high density hydrogen level populations,  we can be much
more  optimistic about
finding this intrinsically interesting  and informative phenomenon   elsewhere
in the Universe.
\ls
\ls
The authors are grateful to D.G. Hummer and P.J. Storey for the   possibility
of using their
machine-readable hydrogen population calculation files prior to their
publication.  V.P. thanks the partial support of this research by the Tomalla
Foundation Fellowship and the International Center for   Fundamental
Physics in Moscow for its help in obtaining this   Fellowship. VSS and VOP
thank the Smithsonian Institution, National Air and Space Museum for   a
senior fellowship and
visiting fellowship, respectively, to work in the Laboratory for Astrophysics
on this program and
related observations.  Together with HAS they acknowledge financial
support from the
Institution's Scholarly Studies Program.  HAS also acknowledges partial
support from NASA grant NAGW-1711.
\vfill\eject
\centerline{REFERENCES}
\vskip 6pt
\ref{Baker, J.G., and Menzel, D.H. 1938, \apj\vol{88} 52.}
\ref{Brockleherst, M., and Seaton, M.J. 1972, \mnras\vol{157} 179.}
\ref{Cilli\'e, G.G. 1936, \mnras\vol{96} 771.}
\ref{Elitzur, M. 1992, {\it Astronomical Masers}, Boston: Kluwer.}
\ref{Gee, C. S., Percival, I. C., Lodge, J. G., and Richards, D.     1976,
\mnras\vol{175} 209.}
\ref{Goldberg, L. 1966, \apj\vol{144} 1225.}
\ref{Gordon, M.A. 1992, \apj\vol{387} 701.}
\ref{Gordon, M.A. 1994, \apj\vol{421} 314.}
\ref{Hummer, D.G., and Storey, P.J. 1992, \mnras\vol{254} 277.}
\ref{Krolik, J.H. and McKee, C.F. 1978, \apjs\vol{37} 459.} \ref{Kronberg,
P.P., Biermann, P., and Schwab, F.R. 1985,   \apj\vol{291} 693.}
\ref{Kurtz, S., Churchwell, E., and Wood, D.O.S. 1993, \apjs\vol{91}   659.}
\ref{Mart\'in--Pintado, J., Bachiller, R., Thum, C., and Walmsley, C.M.
1989a, \aap\vol{215} L13.}
\ref{Mart\'in--Pintado, J., Thum, C., Bachiller, R. 1989b,   \aap\vol{229} L9.}
\ref{Menzel, D.H. 1937, \apj\vol{85} 330.}
\ref{Osterbrock, D.E. 1989, {\it Astrophysics of Gaseous Nebulae and Active
Galactic Nuclei}, Mill Valley, CA: University Sci.Books.}
\ref{Planesas, P., Mart\'in--Pintado, J., and Serabyn, E. 1992, \apjl\vol{386}
L23.}
\ref{Ponomarev, V.O. 1994, \paz\vol{20} 184.}
\ref{Ponomarev, V.O., Smirnov, G.T., Strelnitski, V.S., Chugai, N.N.
1989, \ats\vol{1540} 5.}
\ref{Ponomarev, V.O., Strelnitski, V.S., Chugai, N.N. 1991,   \ats\vol{1545}
37.}
\ref{Ponomarev, V.O., Smith, H.A., and Strelnitski, V.S. 1994,   \apj\vol{424}
976.}
\ref{Reid, M.J. and Moran, J.M. 1981, \anrev\vol{19} 231.}
\ref{Rosseland, S. 1926, \apj\vol{63} 218.}
\ref{Scoville, N.Z., and Norman, C.A. 1988,\apj\vol{332} 163.} \ref{Scoville,
N.Z., and Norman, C.A. 1995, in preparation.} \ref{Shaver, P.A.,
Churchwell, E., and Rots, A.H. 1977, \aap\vol{55}   435.}
\ref{Seaton, M.J. 1959, \mnras\vol{119} 90.}
\ref{Slettebak, A., Collins II, G.W., and Truax, R., 1992,   \apjs\vol{81}
335.}
\ref{Smith, H.A., Larson, H.P., and Fink, U. 1979, \apj\vol{233}     132.}
\ref{Smith, H.A., Strelnitski, V.S., Thum, C., and Mart\'in--Pintado,   J.
1995,
in preparation.}
\ref{Sobolev, A.M., Strelnitski, V.S., and Chugai, N.N. 1985, {\it
Astrofisica~{\rm,}\ }{\bf {22}{\rm,}\ } 613.}
\ref{Storey, P.J., and Hummer, D.G 1995, \mnras\vol{272} 41.}
\ref{Strelnitski, V.S. 1974, {\it Soviet Physics Uspekhi~{\rm,}\   }\vol{17}
507.}
\ref{Strelnitski, V.S. 1983, {\it Nauchnye Informatsii Astron.
Council USSR Ac. Sci.~{\rm,}\ }\vol{52} 75. (Spanish translation:     {\it
Ciencia}, \vol {43}
185.)}
\ref{Strelnitski, V.S. 1984, \mnras\vol{207} 339.}
\ref{Strelnitski, V.S. 1993, In: ``Astrophysical Masers,'' Proceedings of a
Conference Held in
Arlington, Virginia, USA 9-11   March 1992, ed. A.W.~Clegg and
G.E.~Nedoluha;
Springer-Ferlag, Berlin etc., 1993, p.15.}
\ref{Strelnitski, V.S., Smith,  H.A., Haas,  M.R., Colgan, S.W.J.,
Erickson,  E.F., Geis, N., Hollenbach, D.J., and Townes, C.H. 1995a, In:
Proceedings of the
Airborne Astronomy Symposium on the Galactic Ecosystem: From
Gas to Stars to Dust,
ed. M.R. Haas, J.A. Davidson, and E.F. Erickson; San-Francisco: ASP;
p.271.}
\ref{Strelnitski, V.S., Smith, H.A., and Ponomarev, V.O. 1995b, \apj
submitted (``Paper 3'').}
\ref{Thum, C., Mart\'in--Pintado, J., and Bachiller, R. 1992, \aap\vol{256}
507.}
\ref{Thum, C., Matthews, H.E., Mart\'in--Pintado, J., Serabyn, E.,
Planesas, P., and Bachiller, R.
1994a, \aap\vol{283} 582.}
\ref{Thum, C., Matthews, H.E., Harris, A.I., Tacconi, L.J., Shuster,   K.F.,
and Mart\'in--Pintado,
J. 1994b, \aap\vol{288} L25.}
\ref{Thum, C., Strelnitski, V.S., Mart\'in--Pintado, J., Matthews,
H.E., and Smith, H.A. 1995, \aap submitted.}
\ref{Walmsley, C.M. 1990, \aas\vol{82} 201.}
\ref{Wood, D.O.S., and Churchwell, E. 1989, \apjs\vol{69} 831.} \ls
\vfill\eject
\centerline{FIGURE CAPTIONS}

\ls
Fig.~1. To the classification of non-LTE states of a quantum   transition.
\ls
Fig.~2. {\it Solid lines}: $b_n\,(n)\,$ at $T_e=10^4\,$K and at   different
electron number densities $N_e$ (cm$^{-3}$), indicated near   the curves.
Numerical data from Storey and Hummer (1995). {\it Broken   line}:
$b_n\,(n)\,$ at $N_e = 0\,$ (``radiative limit''), calculated   with the
program
of one of us (V.P.) mentioned in text.
\ls
Fig.~3. To the explanation of the overcooling mechanism:   $b_{nl}\,(l)\,$ for
$n = 30 - 45$, at $T_e=10^4\,$K and $N_e =   10^3\,{\rm cm^{-3}}$. Note
the rearrangement of the $b_{nl}\,$   factors in favor of the high-$l$
sublevels with the decrease of $n$,   due to the increasing unblurring of the
$l$-structure. Numerical data   from Storey and Hummer (1995).

\ls
Fig.~4. {\it a} --- \be$_{n,n+1}(n)\,$ at $T_e = 10^4\,$K and at $N_e   = 10^3
- 10^7\,{\rm cm^{-3}}\,$ (numbers near the curves).
\par\noindent
{\it b} --- The same for $N_e = 10^8 - 10^{12}\,{\rm cm^{-3}}\,$.   Numerical
data from Storey and Hummer (1995).
\ls

Fig.~5. \be$_{n,n+1}(N_e)\,$ at $T_e=10^4\,$K for different $n\alpha   \,$
lines ($n\,$ is indicated near the curves). Numerical data from   Storey and
Hummer (1995). The bars, in the lower part of the figure,   show the location
and the width (FWHM) of the $k_l\,(N_e)\,$ and   $k_{net}\,(N_e)\,$ profiles
for the 36\al\ line, according to our   calculations (compare with Fig.~6b).

\ls
Fig.~6. {\it a--i} --- $k_{net}\,(N_e)$ for some \al-lines at $T_e =
10^4\,$K.
{\it j} --- $k_{net}\,(N_e)$ at $T_e = 10^4\,$K for some \be-lines. \ls
Fig.~7. {\it Solide lines}: log\thinspace $|k_{net}|\,({\rm   log}\,N_e)$ for
higher $n$ \al-lines ($n$ is indicated near the   curves), for $T_e = 10^4\,$K.
{\it Broken lines} -- loci of the   maximal $|k_{net}|$ for individual
\al-lines, for
$T_e = 10^4$ and   $5\cdot 10^3\,$K. {\it Dotted line} -- the envelope of the
$|k_{net}|\,(N_e)$ plots for $T_e = 5\cdot 10^3\,$K (the plots are   not
shown).
\ls
Fig.~8 Log\thinspace $|k_{net}|\,(n)$ for different electron   densities, for
$T_e = 10^4\,$K (solide lines) and $5\cdot 10^3\,$K   (broken lines).
\ls
Fig.~9. {\it Dots}: Graphical representation of the
$N_e^{max}\,(n)\,$ dependence, with the numerical data from Table~1   for
$T_e = 10^4\,$K. {\it Dashed line}: approximation (3.1.2) for $5   \le n \le
40$.
\ls
Fig.~10. The LTE line and continuum absorption coefficients, $k_l'\,$   and
$k_c'$, as functions of $n$, for $T_e = 10^4$ and $5\cdot   10^3\,$K.
\ls
\ls
\centerline{TABLE CAPTION}
\ls
Table 1. Optimum densities of unsaturated maser amplification,
$N_e^{max}$, and the values of
$k_{net}^{max}\,$ for hydrogen   recombination \al- and \be-lines at   $T_e
=\,$ 10,000 and 5,000K.
\vfill\eject
VLADIMIR S. STRELNITSKI: Laboratory for Astrophysics, MRC 321,
National Air and Space Museum, Smithsonian Institution, Washington,
DC 20560.
{\it e-mail:} vladimir@wright.nasm.edu
\vskip6pt
HOWARD A. SMITH: Laboratory for Astrophysics, MRC 321,
National Air and Space Museum, Smithsonian Institution, Washington,
DC 20560.
{\it e-mail:} howard@wright.nasm.edu
\vskip6pt
VICTOR O. PONOMAREV: Astro-Space Center of P.N. Lebedev Physical
Institute, Leninsky
prospect, 53, Moscow 117924. {\it e-mail:}
ponomarev@rasfian.serpukhov.su
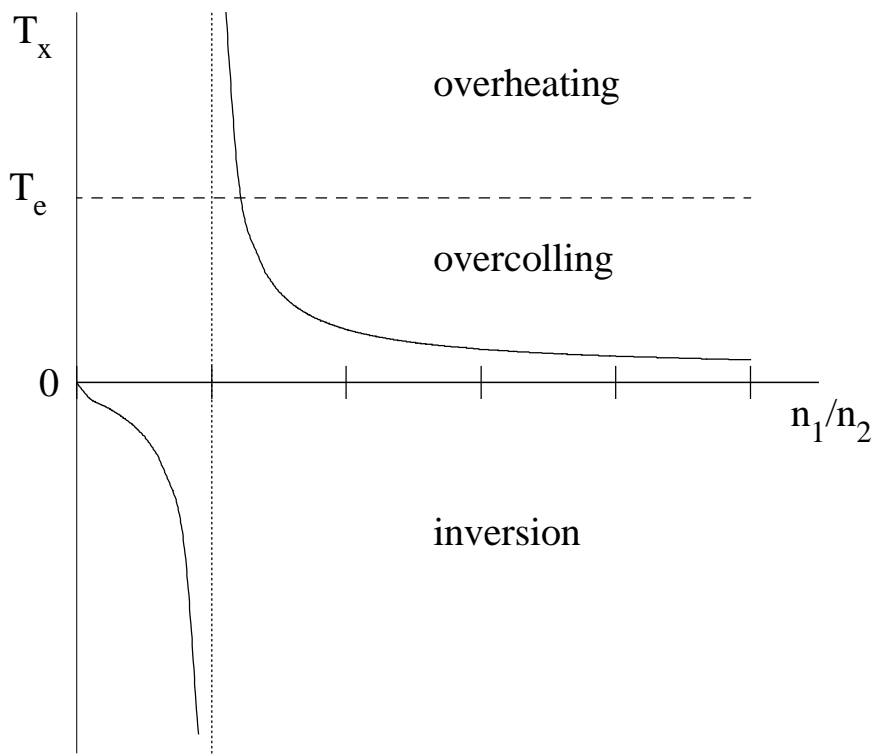
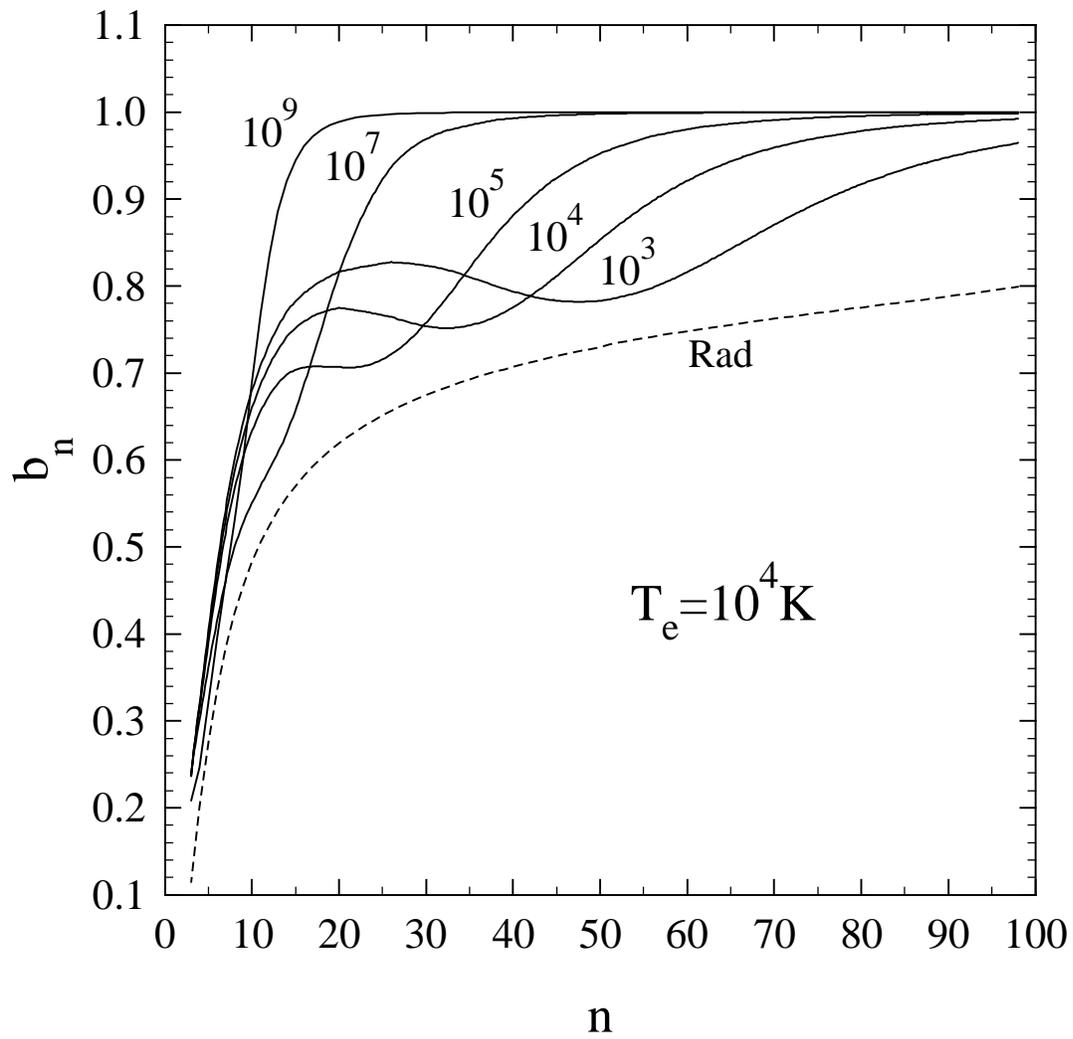
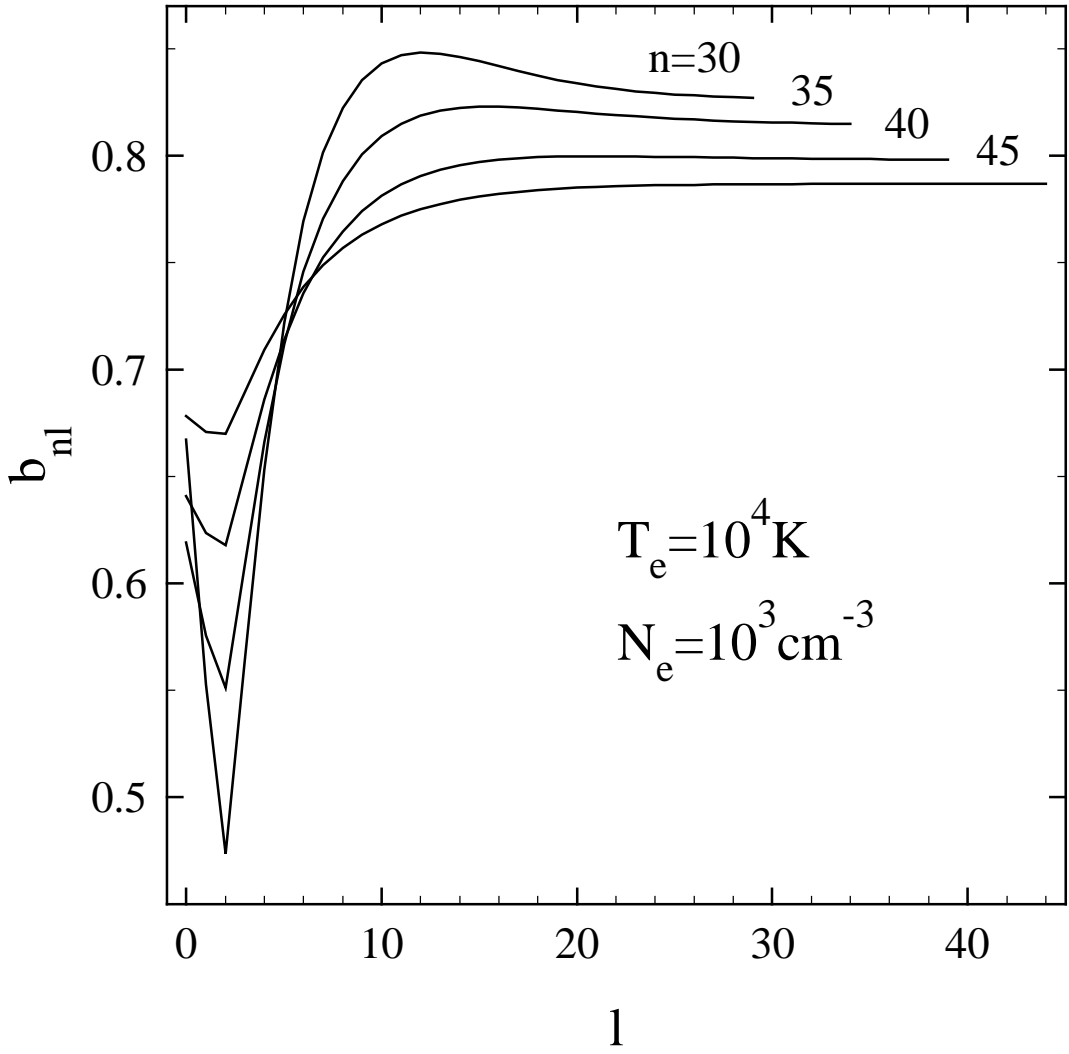
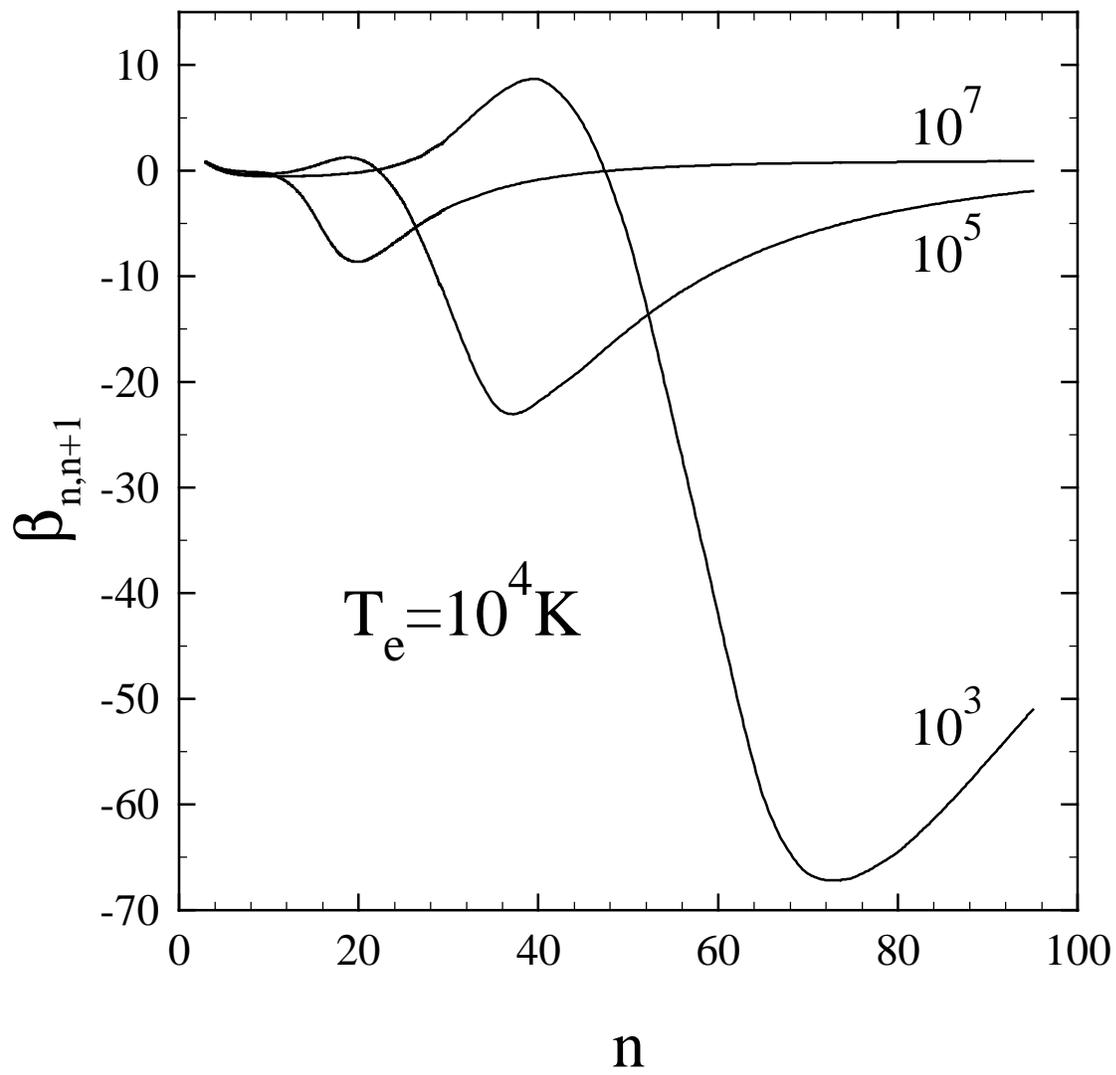
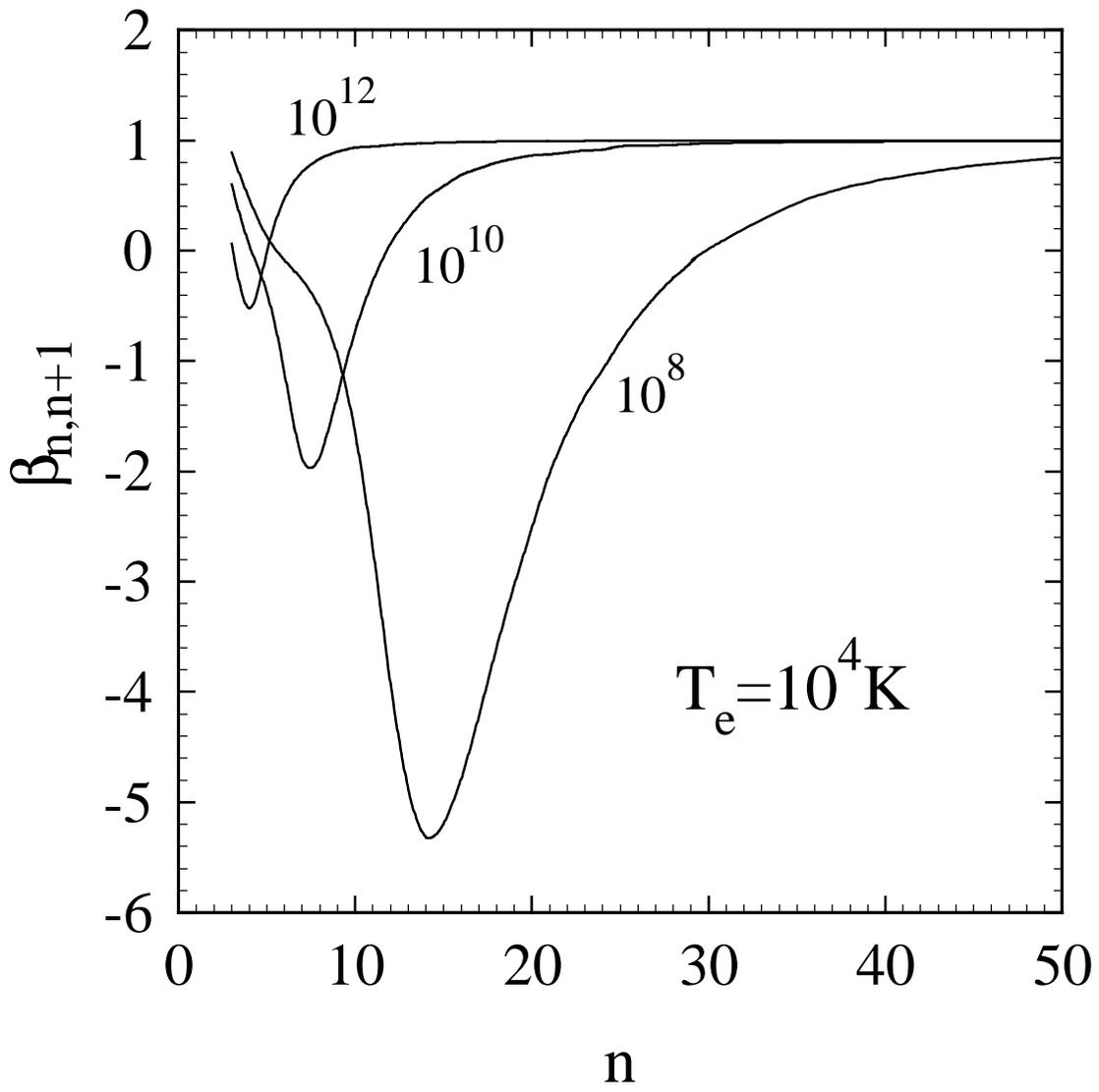
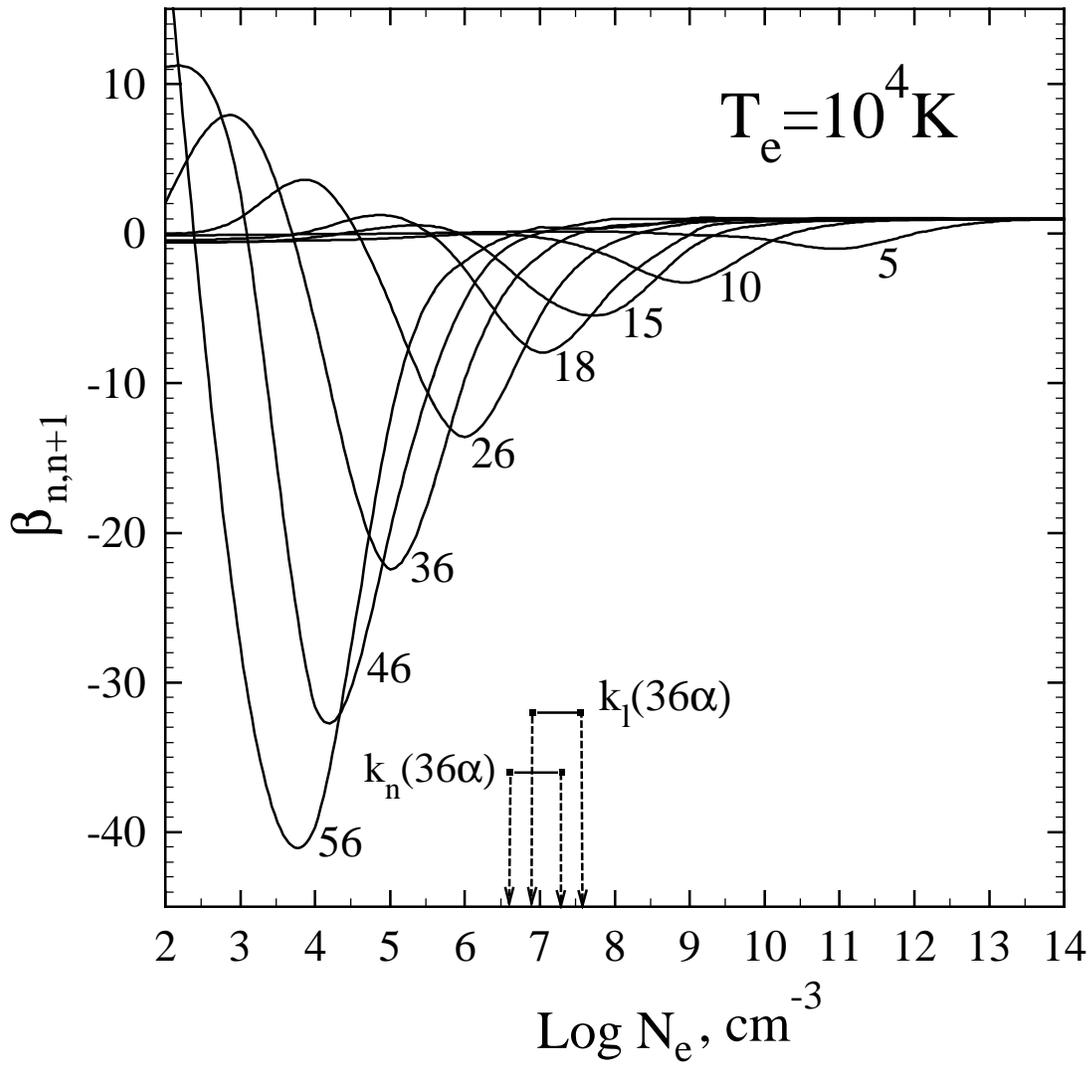
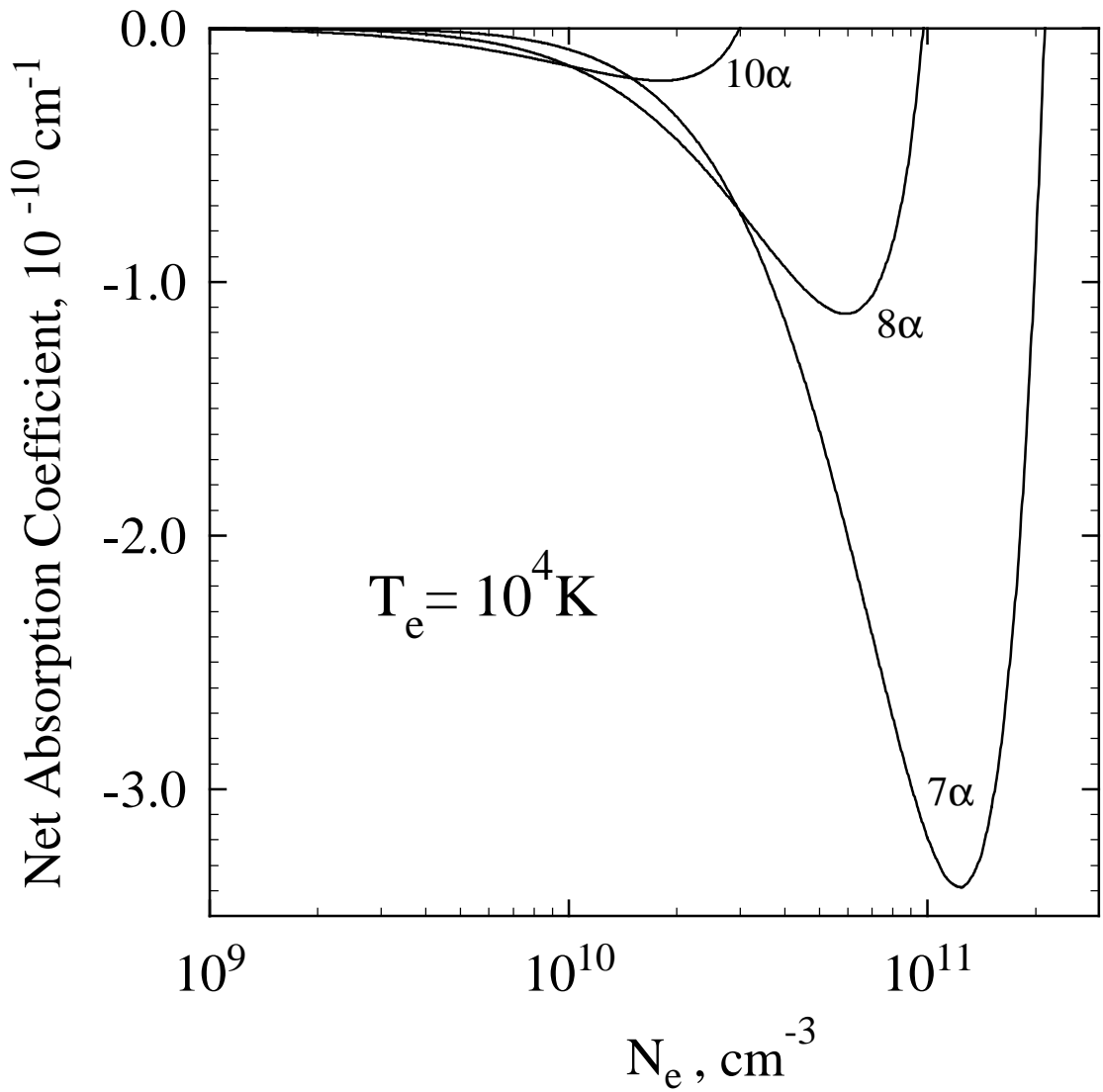
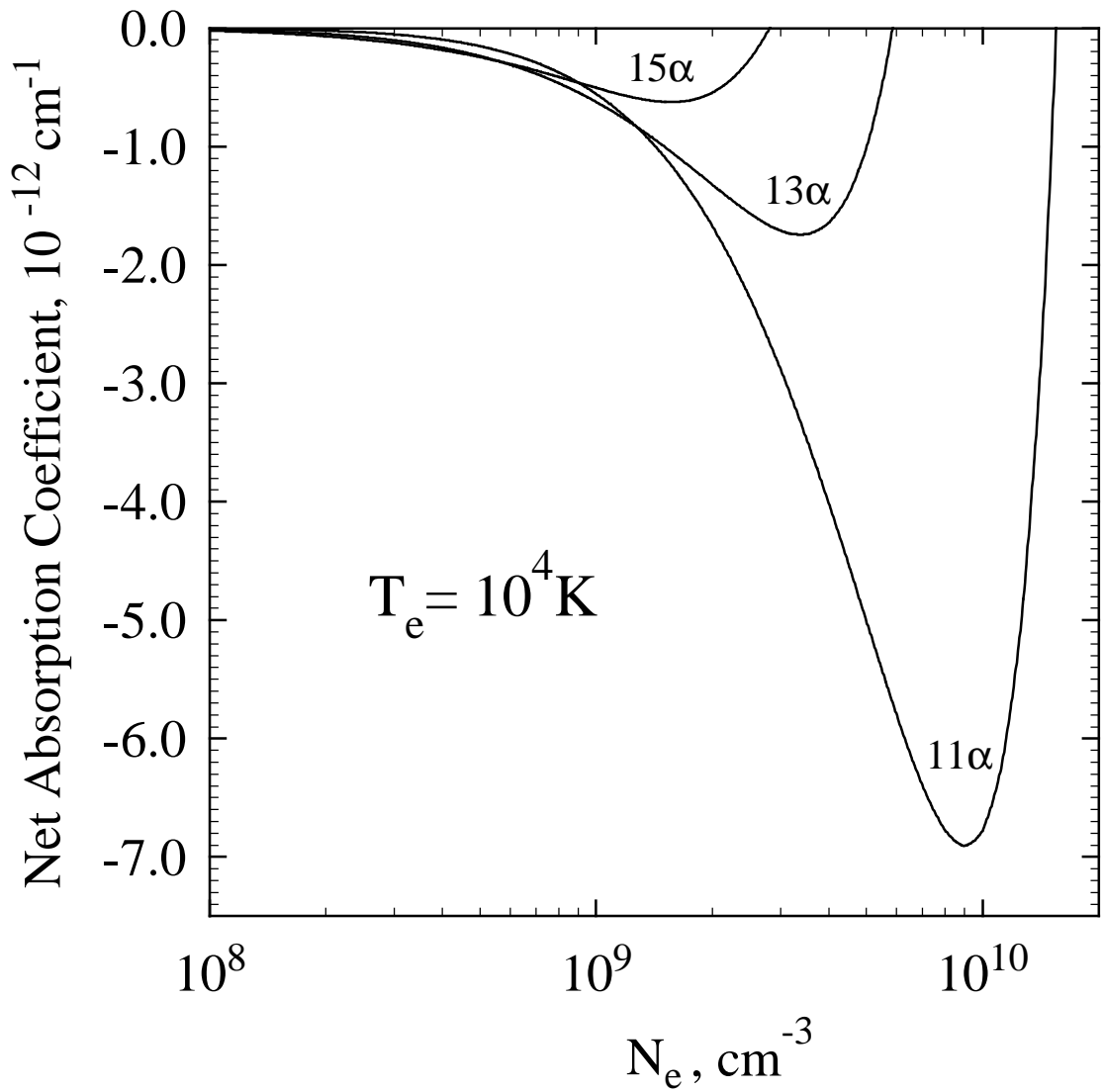
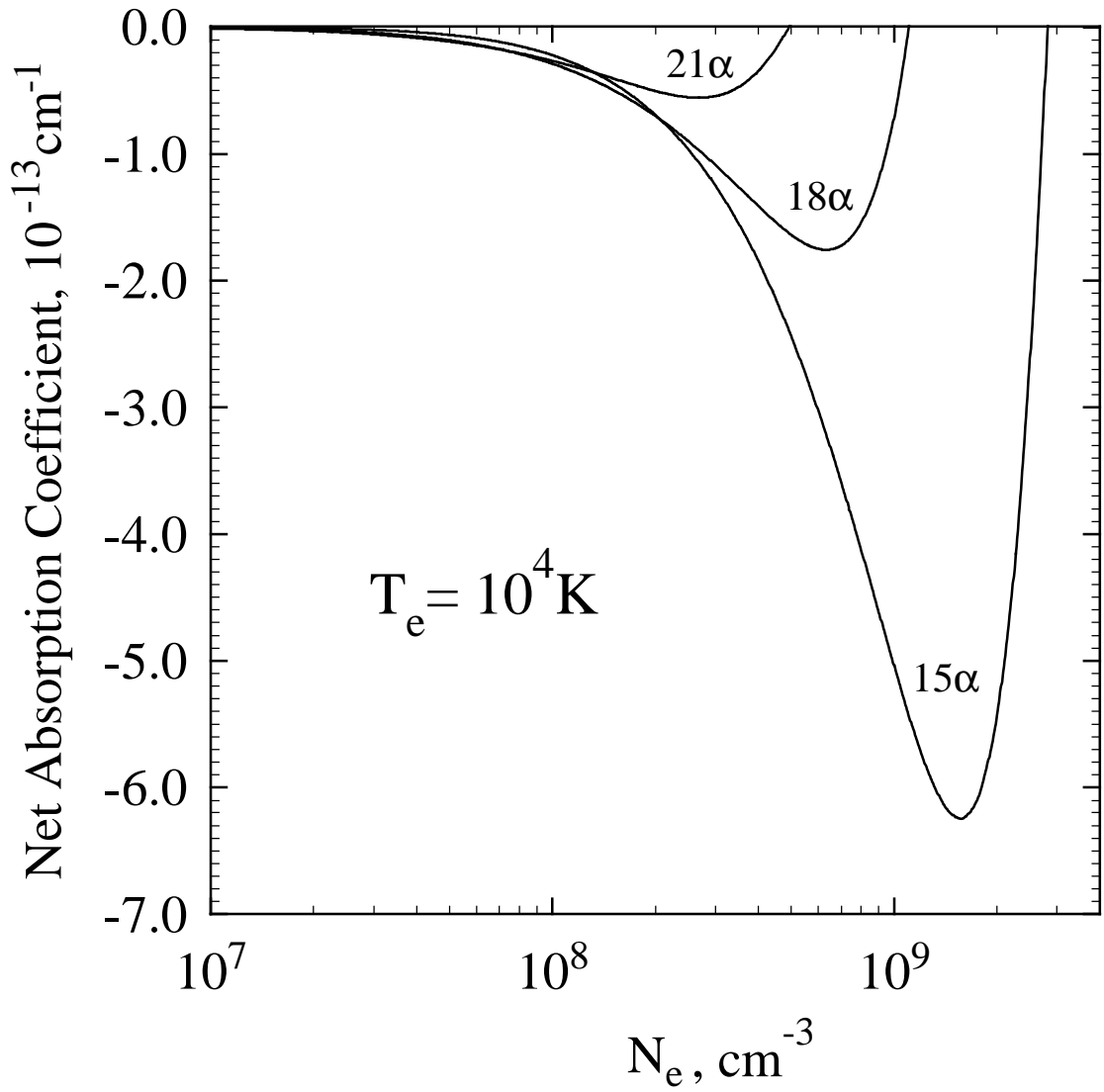
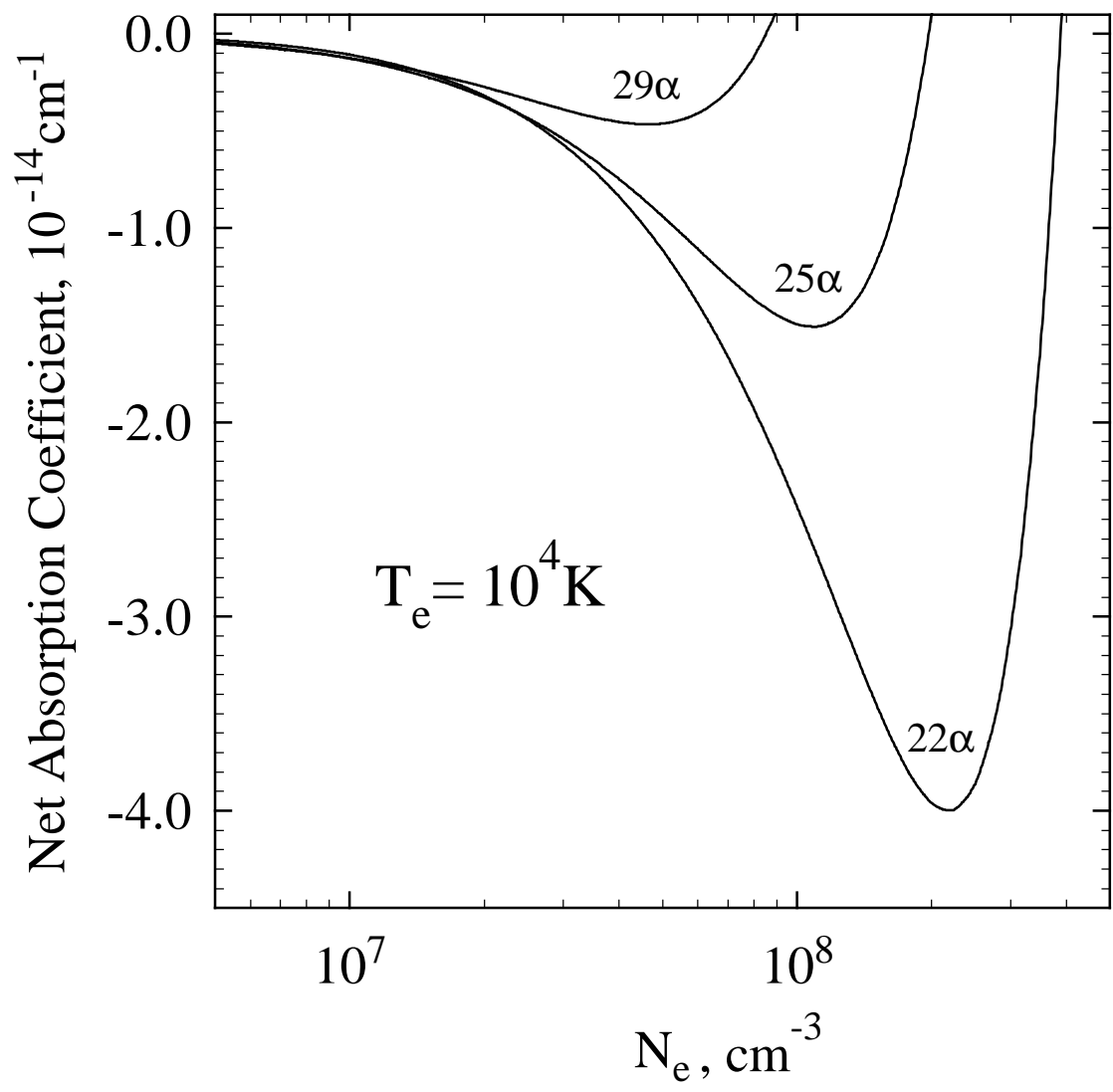
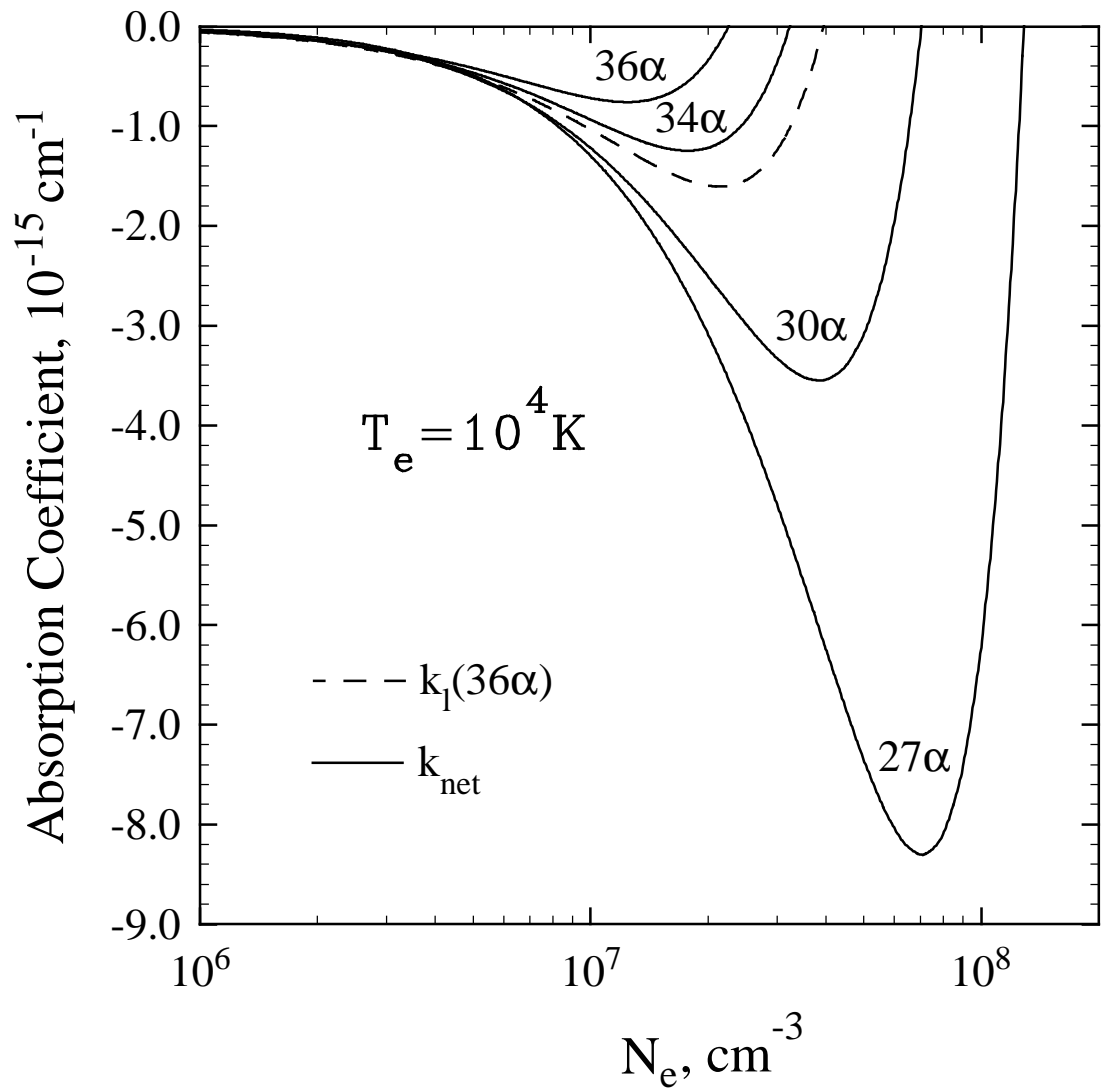
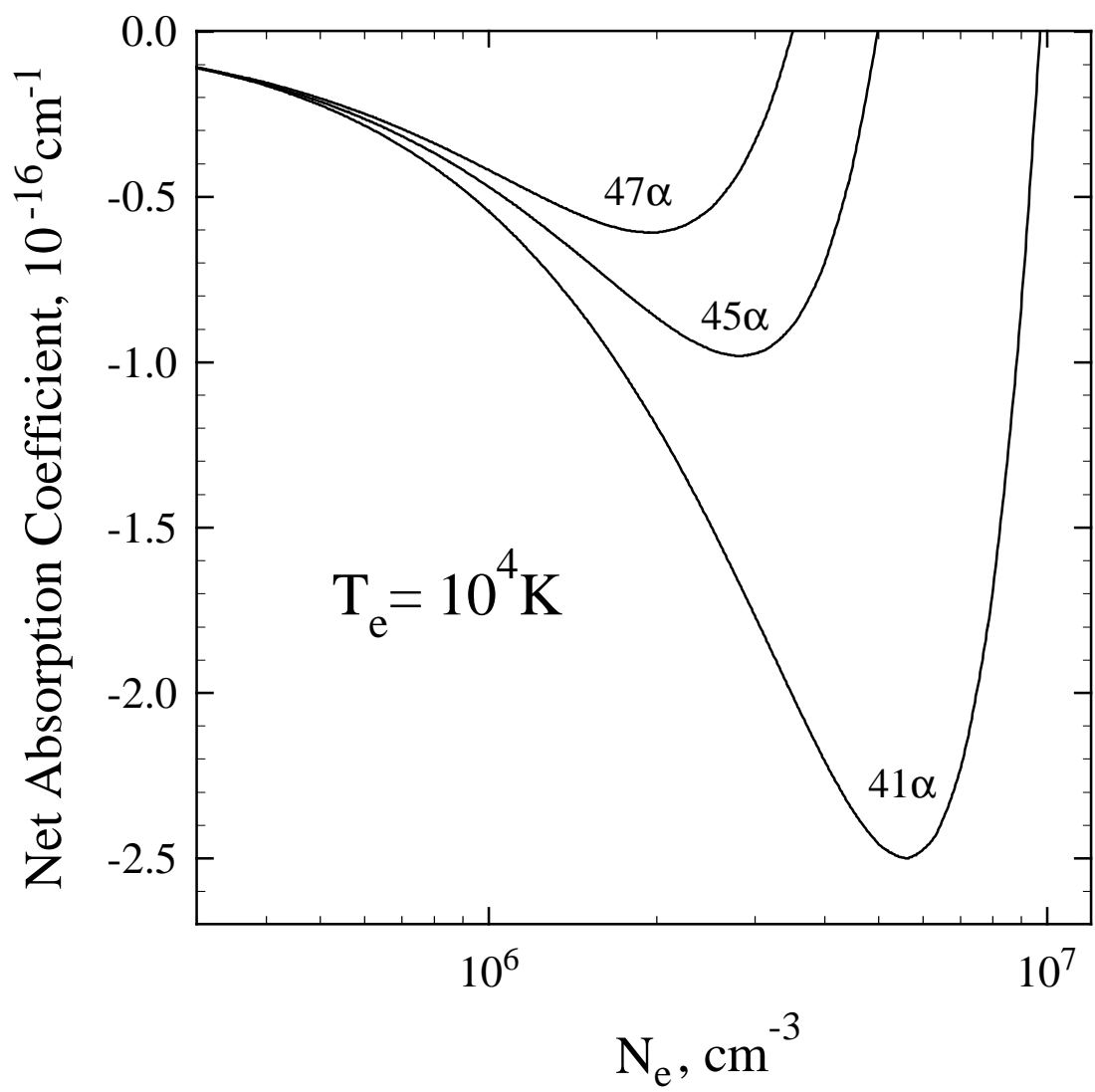
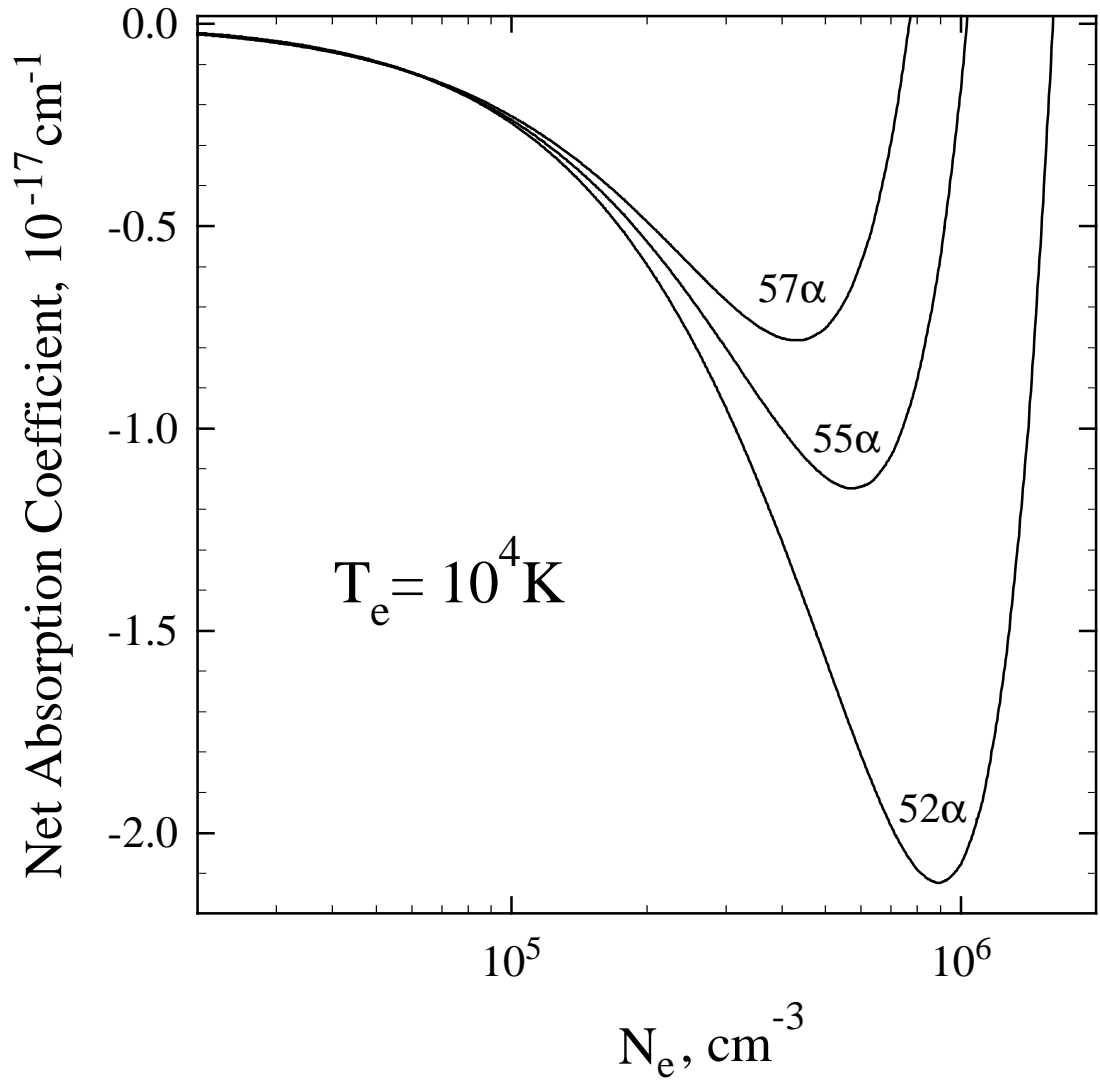
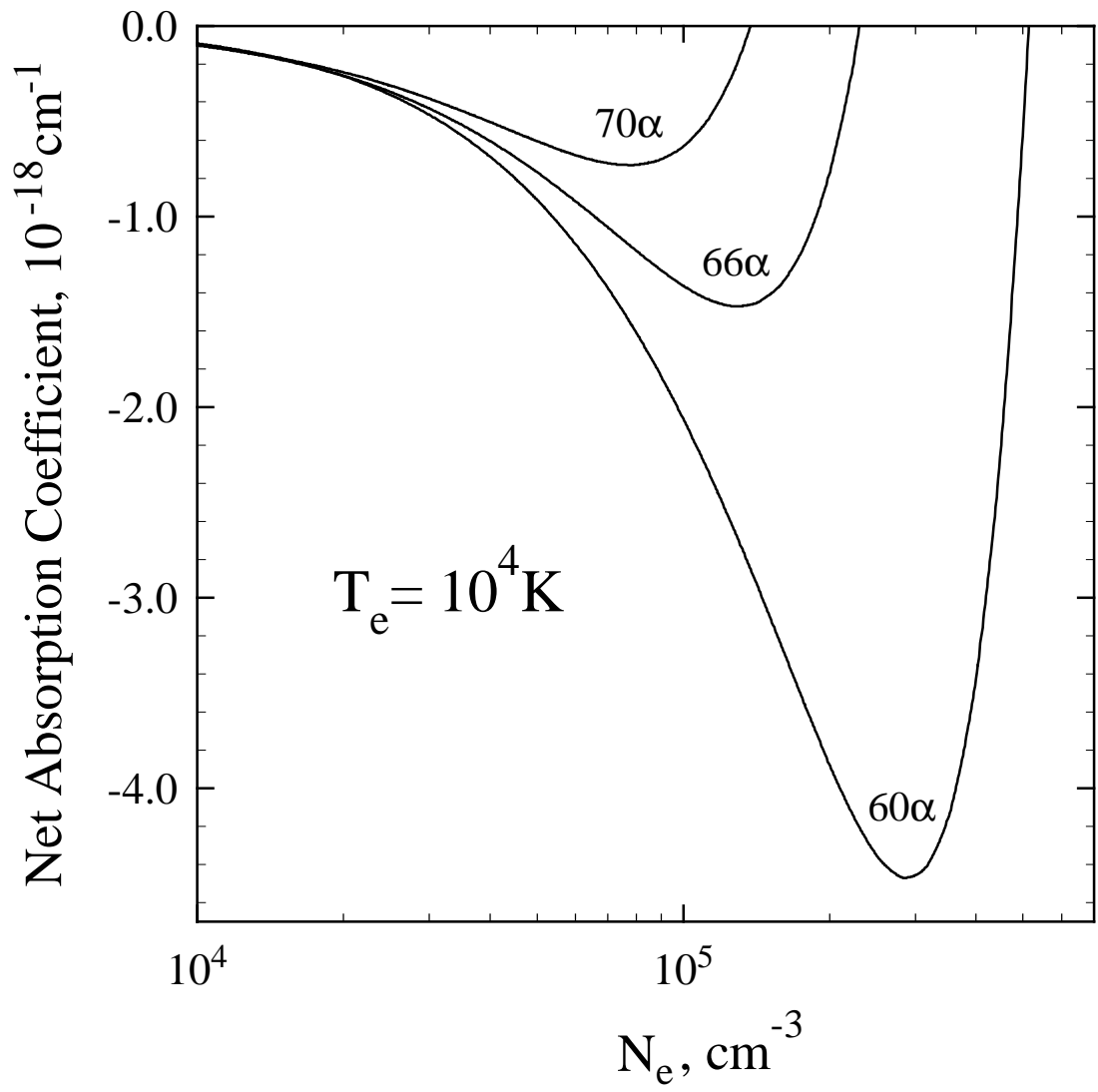
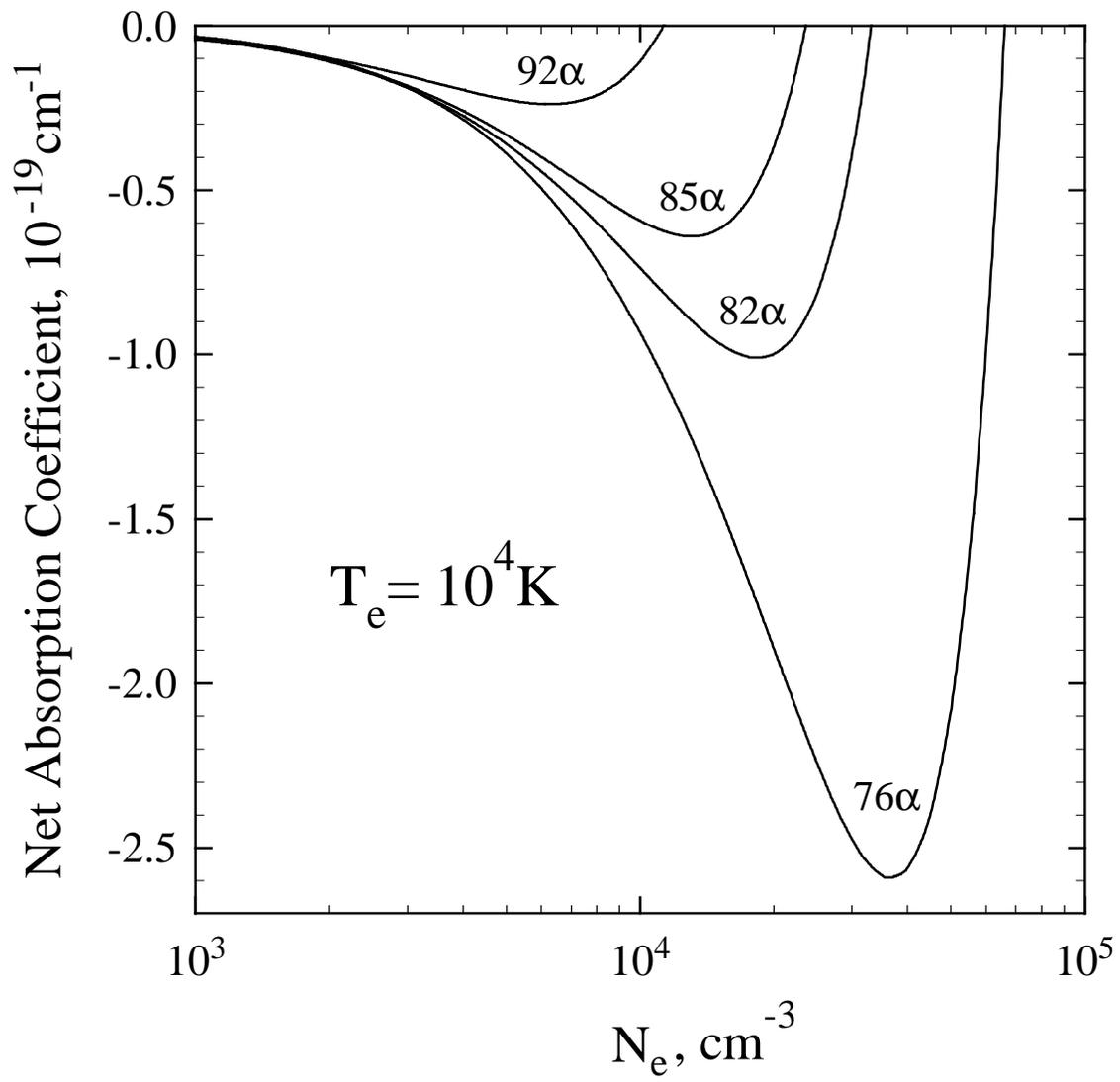
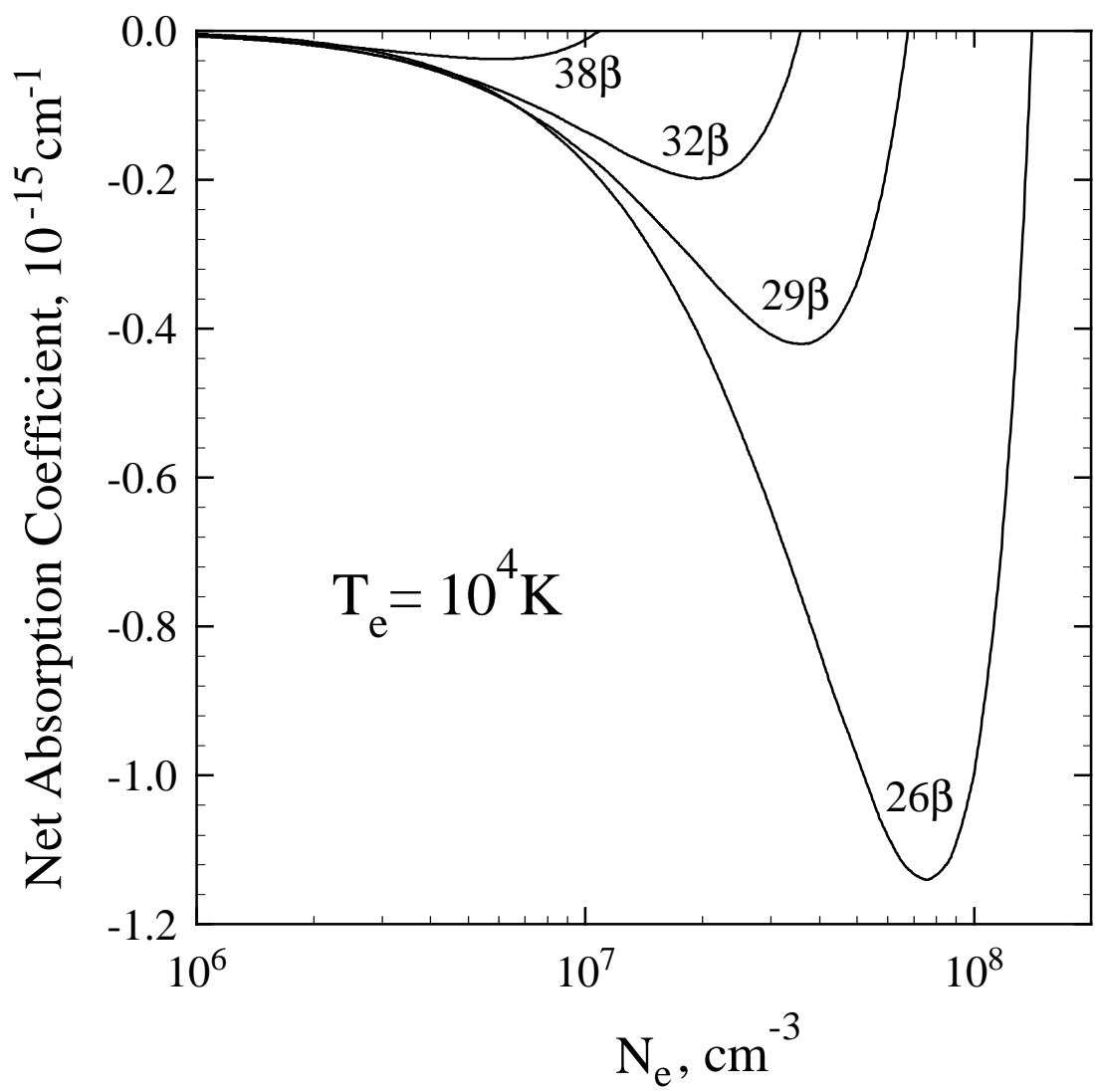
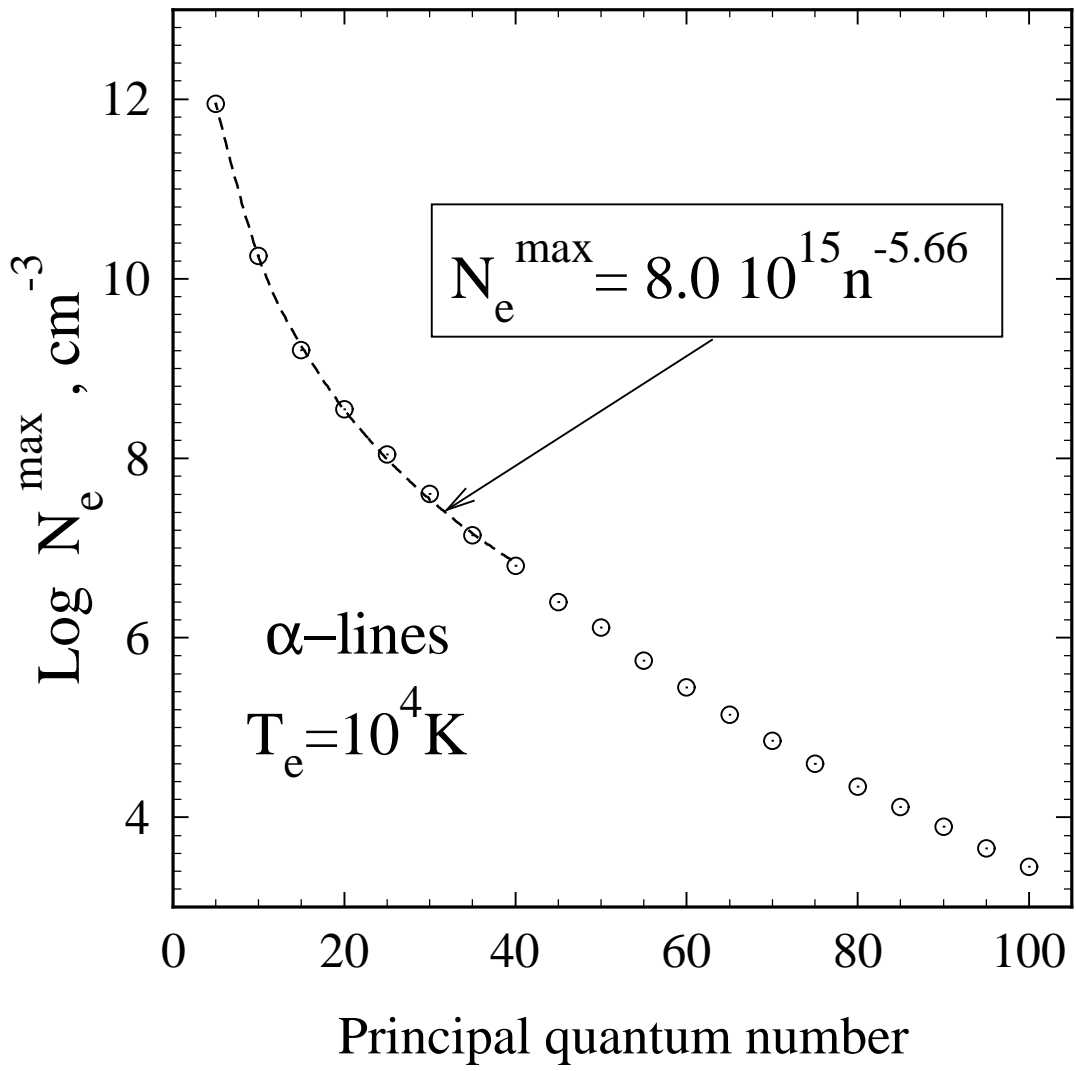
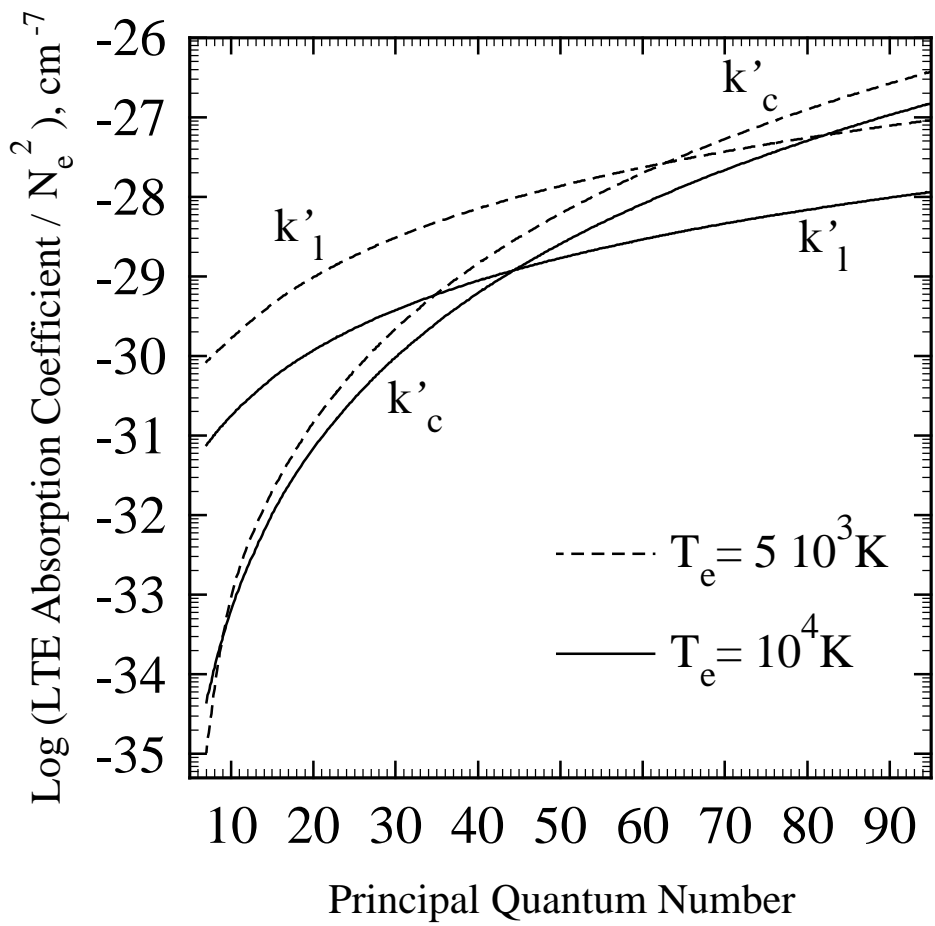
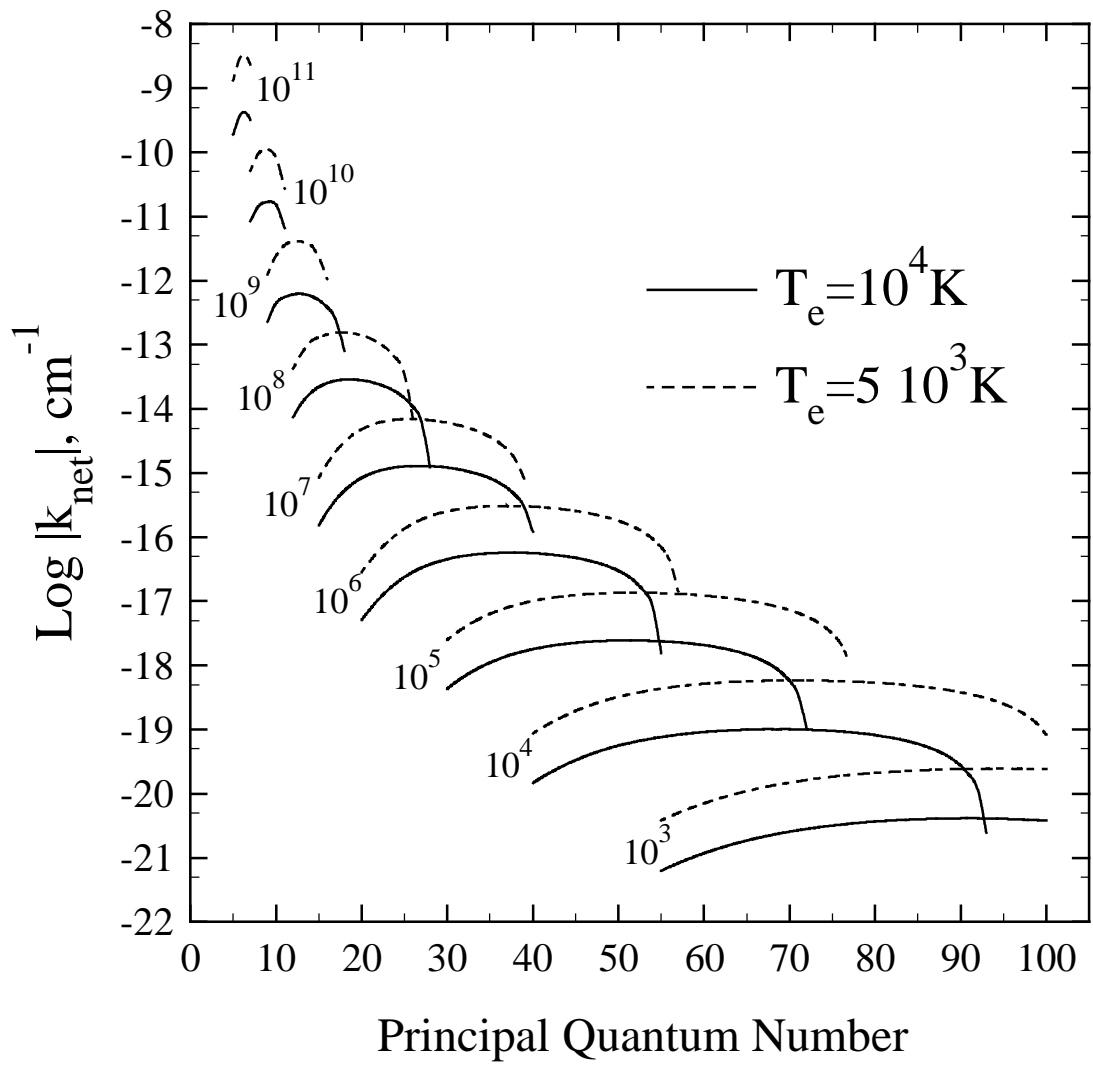
\end